\documentclass{article}
\usepackage{mathrsfs}
\usepackage{a4wide}
\usepackage{amsfonts,amssymb,amsthm,dsfont}
\usepackage{mathtools} 
\usepackage{chemarrow} 
\usepackage{bbold}

\usepackage[active]{srcltx}

\DeclareMathAlphabet{\mathpzc}{OT1}{pzc}{m}{it}
\def \beq{\begin{equation}}
\def \eeq{\end{equation}}

\def\and {{\rm \; and \;}}

\newcommand {\pa}{\partial}

\renewcommand{\Im}{\rm Im\,}
\renewcommand{\Re}{\rm Re\,}

\usepackage{graphicx}
\DeclareGraphicsExtensions{.eps,.bmp}

\newtheorem{theorem}{Theorem}[section]
\newtheorem{thm}{Theorem}[section]
\newtheorem{definition}[theorem]{Definition}
\newtheorem{proposition}[theorem]{Proposition}

\newtheorem{lemma}[theorem]{Lemma}

\theoremstyle{definition}

\newcommand{\eh}{\hfill}
\newlength{\sperrT}

\numberwithin{equation}{section}

\renewcommand{\v}[1]{\mbox{\boldmath$#1$}} 
\renewcommand{\vec}[1]{\v{#1}}
\newcommand{\op}[1]{\mathrm{\mathbf{#1}}} 
\newcommand{\opG}[1]{\mbox{\boldmath{$#1$}}} 

\DeclarePairedDelimiter{\abs}{\lvert}{\rvert} 
\DeclarePairedDelimiter{\norm}{\lVert}{\rVert} 
\DeclarePairedDelimiter{\flux}{\operatorname{fl}(}{)} 
\newcommand{\FL}{\textup{(FL)}}

\DeclarePairedDelimiter{\trace}{\operatorname{Tr}\{}{\}} 
\DeclarePairedDelimiter{\traceN}{\textup{Tr}\{}{\}} 
\newcommand{\traceNtxt}{\operatorname{Tr}}
\newcommand{\traceTxt}{\operatorname{Tr}}

\usepackage{amsfonts}
\newcommand{\field}[1]{\mathbb{#1}} 
\newcommand{\Z}{\field{Z}} 
\newcommand{\N}{\field{N}}

\newcommand{\R}{\field{R}}

\newcommand{\UnulN}[2]{e^{#1 i #2 \op{H}_{\textup{N},\textup{b}}}}

\newcommand{\pdiff}[3][]{\frac{\partial^{#1} #2}{\partial #3^{#1}}} 
\newcommand{\diff}[3][]{\frac{\textup{d}^{#1} #2}{\textup{d} #3^{#1}}} 

\newcommand{\identity}{\op{I}}

\renewcommand{\Re}{\textup{Re}}
\renewcommand{\Im}{\textup{Im}}

\newcommand{\diffd}{\textup{d}}

\newcommand{\subN}{\textup{N}}
\newcommand{\subb}{\textup{b}}
\newcommand{\subFD}{\textup{FD}}
\newcommand{\subE}{\textup{E}}

\newcommand{\subF}{\textup{F}}

\newcommand{\kv}[1]{(#1)} 
\newcommand{\gG}{\Gamma}    
\renewcommand{\gg}{\gamma} 
\newcommand{\vg}{\v{\gg}} 
\newcommand{\gR}{\Gamma^*}    
\newcommand{\gr}{\gamma^*} 
\newcommand{\vr}{\v{\gr}} 
\newcommand{\gT}{\Lambda}    
\newcommand{\gTN}{\Lambda_{N}}    
\newcommand{\gTNt}{\tilde{\Lambda}_{N}} 
\newcommand{\gTNtt}{\tilde{\tilde{\Lambda}}_{N}} 
\newcommand{\gt}{x}    
\newcommand{\gtt}{y}
\newcommand{\vt}{\v{\gt}}    
\newcommand{\vtt}{\v{\gtt}}
\newcommand{\gUC}{\Omega}     
\newcommand{\gu}{\underline{\gt}}           
\newcommand{\vu}{\v{\gu}}           
\newcommand{\gut}{\underline{\gtt}}           
\newcommand{\vut}{\v{\gut}}           
\newcommand{\gBZ}{\Omega^*}   

\newcommand{\chN}{\chi_{\subN}}
\newcommand{\chNt}{\tilde{\chi}_{\subN}}
\newcommand{\chNtt}{\tilde{\tilde{\chi}}_{\subN}}

\newcommand{\sz}{\mathcal{S}}
\newcommand{\szm}{\mathcal{S}_{-}}

\newcommand{\sk}{\mathcal{Q}}
\newcommand{\skm}{\mathcal{Q}_{-}}

\newcommand{\hlbG}{\mathscr{H}} 

\begin{document}

\noindent 
\begin{center}
\textbf{\large On the Verdet constant and Faraday rotation for graphene-like materials }
\end{center}

\begin{center}
February 19, 2013
\end{center}

\vspace{0.5cm}

\begin{center}
\textbf{
Mikkel H. Brynildsen\footnote{Department of Mathematical Sciences,
  Aalborg University, Fredrik Bajers Vej 7G, 9220 Aalborg, Denmark},
Horia D. Cornean\footnotemark[\value{footnote}]
}

\end{center}

\begin{abstract}
We present a rigorous and rather self-contained analysis of the
Verdet constant in graphene-like materials. We apply the 
gauge-invariant magnetic perturbation theory to a nearest-neighbour tight-binding model 
and obtain a
relatively simple and exactly computable formula for the Verdet constant, at all temperatures
and all frequencies of sufficiently large absolute value. Moreover, for the standard nearest neighbour tight-binding model
of graphene we show that the transverse component of the conductivity
tensor has an asymptotic Taylor expansion in the external magnetic
field where all the coefficients of even powers are zero. 
\end{abstract}

\noindent

\section{Introduction}

Faraday rotation is a dispersion effect discovered around 1845, which consists of the rotation of
the polarization plane of a linearly polarized light-beam passing through a material in the
direction of the applied magnetic field. The  Faraday rotation angle 
$\vartheta$ is defined as
\[
\vartheta = \frac{\omega d(\eta_{-} - \eta_{+})}{2c}, 
\]
where $d$ is the thickness of the material, while $\eta_{-}$ and
$\eta_{+}$ are respectively the refraction indices of the 
right and left circularly polarized radiation of frequency
$\omega$.
As we will explain in Section \ref{FaradaySigma}, if the applied magnetic
field $b$ is small, then the rotation angle can be expressed as
$\vartheta = d\,b\,V$ where $V$ is a constant named after
Emile Verdet \cite{Verdet}, one of the first physicists who advocated the use of
Maxwell's equations in explaining dispersion (around 1865). The Verdet constant depends on how the transverse conductivity
coefficient behaves as a function of $b$ near $b=0$. 

In this paper we put to work the general method proposed in
\cite{FaradayRevisited} and apply it to the case of a tight-binding
model. One advantage of a discrete model is that we can give much
shorter and less technical proofs for both the thermodynamic and adiabatic
limits. Another advantage is that the structure of the Bloch bands is
much simpler in the discrete case, where in many interesting cases we
know the exact expression of the fiber Hamiltonians, given by finite
dimensional matrices. 

One can relax the decay properties of the zero-field Hamiltonian, 
but the nearest neighbour tight-binding operator is widely used by physicists. 
The proofs could be carried out even for a $\op{H}_0$ with a sufficiently fast polynomial decay around the diagonal, without changing the results, 
but the price would be a complication of the thermodynamic limit.

\subsection{A description of the Faraday effect}
Let us briefly discuss the physical problem following  
\cite{Jackson} and \cite{PhysRevB.44.4021}. Starting from the classical Maxwell
equations, one can derive the Faraday rotation angle for a two-dimensional quasi-free electron
gas placed in a magnetic field $B$ (of strength $b$) applied
perpendicular to the layer. The incident light-beam is monochromatic with frequency $\omega$. 
The electrons are confined within a slab of thickness $d$ and move freely along the $xy$
plane. 
Denoting by $\left\{\sigma_{jk}^{(3\textup{D})}\right\}_{j,k=1}^{3}$
the three-dimensional conductivity tensor of the slab, and by
$\sigma_{\pm}^{(3\textup{D})} = \sigma_{11}^{(3\textup{D})} \pm i
\sigma_{21}^{(3\textup{D})}$, the complex refraction indexes $(\eta
- i \kappa)$ for the right $(+)$ and left $(-)$ circularly polarized light are given by
\begin{equation}
(\eta_{\pm}-i\kappa_{\pm})^2 = \mu \epsilon\left[
 1 - (4\pi i/\omega \epsilon) \sigma_{\pm}^{(3\textup{D})}
\right].\label{etapm001}
\end{equation}
The element $\sigma_{33}^{(3\textup{D})}$ does not contribute to
the Faraday rotation. Considering a finite thickness $d$ of
the slab, the three-dimensional conductivity tensor is related to the
two-dimensional conductivity tensor by the following expression
\[
\sigma^{(2\textup{D})}_{jk} =  \sigma^{(3\textup{D})}_{jk}\,d, \qquad j,k \in \{1,2\}. 
\]
Let us define the real-valued function %
\[
f(u,v) =
\left[
 \left( 1 + \frac{4\pi}{d \omega\epsilon} u \right) ^2
+
\left(
\frac{4\pi}{d \omega
    \epsilon}v
\right)^2
\right]^{\frac{1}{4}}
\cos\left(\frac{1}{2}
\arctan \left[  
\frac{ v }{ \left(\frac{d \omega \epsilon}{4 \pi} +u\right)}
\right] 
\right), 
\] 
where $\abs{u} < \frac{ d \omega \epsilon}{4 \pi}, v \in \R$. Now we can
express $\eta_{\pm}$ 
appearing in expression~\eqref{etapm001} (to be chosen non-negative from physical considerations) as: 
\[
\eta_{\pm} = \sqrt{\mu \epsilon}\, f\left(\pm \sigma^{(2\textup{D})}_{21}, \sigma^{(2\textup{D})}_{11}\right).  
\]
We are interested in studying how the Faraday angle $\vartheta$
behaves as a function of the strength of 
the external magnetic field, $b$.
Let us assume (we will prove it later) that one can write down an asymptotic expansion of
$\sigma_{jk}^{(2\textup{D})}(b)$ in powers of $b$. Moreover we assume
that for the off-diagonal tensor element 
we have: 
\[
\sigma_{21}^{(2\textup{D})}(b) = b\sigma_{21}^{(1)}+ \mathcal{O}(b^2)\, ,
\]
while $\sigma_{11}^{(2\textup{D})}(b)$ can be expressed as:
\[
\sigma_{11}^{(2\textup{D})}(b) =\sigma_{11}^{(0)}+ b\sigma_{11}^{(1)}+
\mathcal{O}(b^2).
\]
Then
\begin{equation}
  \vartheta(b) = -d\,b\,\frac{\omega \sigma_{21}^{(1)} }{c} \pdiff{f}{u} \left(0,\sigma_{11}^{(0)}\right) + \mathcal{O}(b^2). \label{FaradaySigma} 
\end{equation}
Thus the Faraday rotation angle is linear in $b$ near zero, and the linear term
can be put in the form $\vartheta = d\,b\,V$, where $V$ is the Verdet constant.
Calculating the Verdet constant is therefore a question of
finding a computable formula for the coefficient $\sigma_{21}^{(1)}$ in the
asymptotic expansion of the off-diagonal 
conductivity element $\sigma_{21}^{(2\textup{D})}$.

The structure of our paper is as follows. In the rest of this
section we introduce the mathematical notions which are necessary
in order to properly define the transverse conductivity element in
\eqref{eq:J2}. In Section \ref{unu0} we formulate the main result of the paper, namely
Theorem \ref{thm:main}. Sections \ref{unu1}, \ref{unu2} and \ref{unu3}
contain the proofs of the three statements of our main theorem. In
Section \ref{unu4} we present our conclusions. 
%
%
\subsection{The configuration space}
\label{sec:Setup}
\begin{figure}[h!tbp]
\begin{center}
\includegraphics[width=\textwidth]{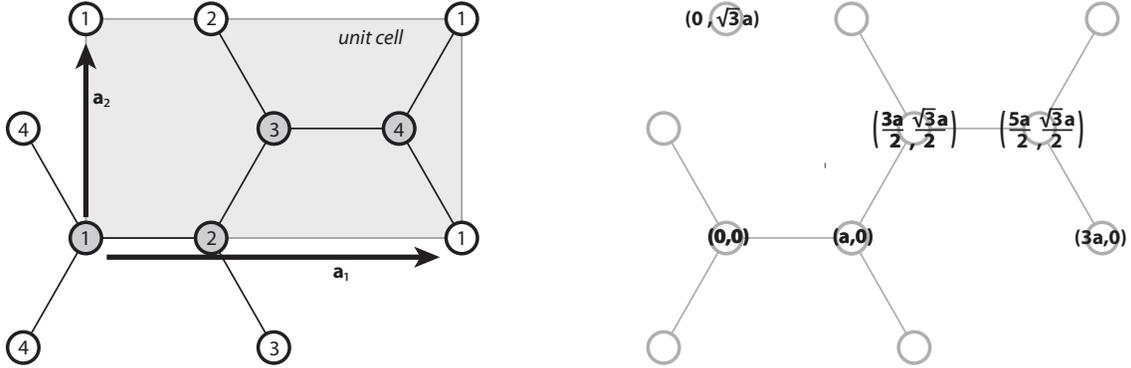}
\caption{\footnotesize{Left: A nearest neighbour tight-binding model of graphene. 
Each unit cell corresponds to a basis of four sites numbered one through four, the grey circles indicate the sites in the unit cell positioned at origin. 
The full lines indicates the non-zero matrix elements $h_0(\vt,\vtt)$, where at least one
of $\vt$ or $\vtt$ belongs to the basis associated with  origin. We denote the nearest-neighbour distance by $a$. The Bravais lattice is generated by the vectors $\v{a}_1$ and $\v{a}_2$. Right: the coordinates of the sites used in the calculations.}}
\label{fig:GrapheneModel}
\end{center}
\end{figure}

We neglect the electron-electron interactions.
To simplify notation we work in a system of units where 
$\hbar = 2m_{\rm electron} = e = 1$.
Let $\v{a}_1 = (a_1,0)$ and $\v{a}_2=(0,a_2)$ be two vectors in $\R^2$. Define the Bravais lattice, 
\[
\gG
= \{ \vg \in \R^2 \; : \; \vg =  m\v{a}_1 + n\v{a}_2, \quad m, n
\in \Z \}.
\]
The basis $\gUC$ is modeled by $\nu_\Omega$ points whose position vectors
are  denoted by $ \{ \vu_n \}_{n=1}^{\nu_{\Omega}}$; one of them is always the
origin, thus also belongs to $\gG$. For graphene $\nu_\Omega=4$ if we want to
have orthogonal generating vectors of the Bravais 
lattice. We denote by $\abs{\gUC} = \abs{a_1 a_2}$ the area of the unit cell
(see figure \ref{fig:GrapheneModel}). 

The total configuration space is  denoted by $\gT$ and describes the set of
ion-core positions. A point in $\gT$ is called a site. Moreover, we have the identity:
\[
\gT = \gG + \gUC=\{ \vg +\vu\in \R^2 \; : \;  \vg\in\gG,\; \vu \in \gUC \}.
\]
Define the reciprocal primitive vectors $\{\v{b}_{n}\}_{n=1}^{\nu_\Omega}$ as:
\[
 \v{a}_i \cdot \v{b}_j = 2 \pi \delta_{ij},
\]
where $\delta_{ij}$ is the Kronecker delta.
The reciprocal lattice (dual lattice) $\gR$ is defined by
\[
\gR = \{\vr \in \R^2 \; : \; \vr = m \v{b}_1 + n\v{b}_2, \quad m \in \Z,n \in \Z \}
\]
We denote by $\gBZ$ the first Brillouin zone for the dual lattice:
\[
\gBZ = \left\{  
t_1\v{b}_1 + t_2\v{b}_2 \, : \, -\frac{1}{2} \leq t_i \leq \frac{1}{2}
\right\}.
\]
The one-electron Hilbert space is $\ell^2(\gT)$. In the absence of the
magnetic field, each electron will be
described by a one-particle Hamiltonian $\op{H}_0$, where
$\op{H}_0$ is a nearest-neighbour tight-binding operator. If we denote
by $\{\delta_{\vt}\}_{{\vt}\in \gT}$ the canonical basis of 
$\ell^2(\gT)$, then $\op{H}_0$ has a kernel
\begin{equation}  
h_0\kv{\vt,\vtt} =\langle \delta_{\vt},\op{H}_0\delta_{\vtt}\rangle
\label{hamilfiber001}
\end{equation}
which is zero if $||\vt-\vtt||$ is larger than some constant. 

Graphene is one material which can be described in this way. Since it
is important to have a 'straight' Bravais lattice, we need to extend
the minimal model, see \cite{wallace1947band}, to a basis with four sites instead of two.
Let $a>0$. We let the basis consist of four atoms $\gUC_G := \left\{ \vu^{(i)} \right\}_{i=1}^{4}$
placed at positions 
\begin{equation}
 \vu^{(1)}= (0,0),\quad 
\quad \vu^{(2)}= (a,0),\quad \vu^{(3)}=  \left(\frac{3a}{2},\frac{\sqrt{3}a}{2}\right)
,\quad \vu^{(4)}=  \left(\frac{5a}{2},\frac{\sqrt{3}a}{2}\right)  ,
\label{basisVectors}
\end{equation}

as indicated in figure \ref{fig:GrapheneModel}. This standard nearest neighbour tight-binding model of graphene 
can be traced back to P. Wallace in his $1947$ article \cite{wallace1947band}. 
The Bravais lattice associated with this model is generated by the two vectors $\v{a}_1=(3a,0)$ and $\v{a}_2=(0,\sqrt{3} a)$, 
see figure \ref{fig:GrapheneModel} , and the first Brillouin zone is thus 
\begin{equation}
\gBZ_G = \left[
-\frac{\pi}{3a},\frac{\pi}{3a}
\right]\times\left[
-\frac{\pi}{\sqrt{3}a},\frac{\pi}{\sqrt{3}a}
\right].
\end{equation}
 In this model a site have
nearest neighbours either at the three relative positions 
\begin{equation}
  (a,0), 
  \quad \left(-\frac{a}{2},\frac{\sqrt{3}a}{2}\right)
  \quad \textup{and} \quad \left(-\frac{a}{2},-\frac{\sqrt{3}a}{2}\right),
\end{equation}
or at the three relative positions $(-a,0),
\left(\frac{a}{2},\frac{\sqrt{3}a}{2}\right)$ and
$\left(\frac{a}{2},-\frac{\sqrt{3}a}{2}\right)$.

 This nearest neighbour tight-binding model leads to a kernel $ h_{0}^G(\vt, \vtt)$ which equals (here
 $\delta$ means Kronecker delta):
\begin{equation}\label{decemb10}
\sum_{k=1}^2\delta{ \left(\gt_1 -\gtt_1, (-1)^k\frac{a}{2}\right) }
  \sum_{j=1}^2\delta{ \left(\gt_2 -\gtt_2, (-1)^j\frac{\sqrt{3}a}{2}\right) }
  + \sum_{l=1}^2\delta{ \left(\gt_1 -\gtt_1, (-1)^l a\right) } \delta{ \left(\gt_2, \gtt_2\right) }. 
\end{equation}
In other words, $ h_{0}^G(\vt, \vtt)=1$ if $||\vt-\vtt||=a$, and $
h_{0}^G(\vt, \vtt)=0$ otherwise.

If we restrict ourselves to a finite crystal, it will be modeled by
\[
\gTN:= \{ (\vg + \vu) \in \gT\; : \;
\vg = m \v{a}_1 + n \v{a}_2, \; \abs{m}\leq N, \abs{n} \leq N
, \vu \in \gUC \}, \qquad N\geq 1.
\]
We denote by $\abs{\gTN}= (2N+1)\abs{\gUC}$ the area of the $2N+1$ unit cells 
covered by $\gTN$.
The characteristic function of the central region is denoted by $\chN$.
The Hamilton operator subject to
Dirichlet boundary conditions (DBC) in $\gTN$ is:
\begin{equation}
\op{H}_{0,\subN} = \chN \op{H}_0 \,\chN. \label{DBC}
\end{equation}
In general, if some operator $\op{O}$ initially defined on 
$\gT$ is afterwards restricted to $\gTN$, we denote this restriction by 
$\op{O}_{\subN} = \chN \op{O} \chN$. 

Now we include a constant, static and external magnetic field into the model, which is thought of as
having always existed (before the light perturbation is turned
on). The magnetic field is assumed to be perpendicular to the layer, directed in the
$z+$ direction, having constant magnitude $b$. 
All our vectors can be seen as three dimensional, and we associate with the symbol $\vt$ both the $2d$ vector $(\gt_1,\gt_2)$ and the $3d$ vector $(\gt_1, \gt_2, 0)$. 
This enables us to use the cross-product shorthand $\vt \times \vtt$ for $2d$ vectors.

Invoking Peierls substitution, our magnetic Hamiltonian
$\op{H}_{\subb}$ in a constant magnetic field
has an integral kernel $h_{\subb}(\vt,\vtt)$ which is obtained by multiplying $h_0(\vt,\vtt)$ by a 
phase factor~\cite{HoriaHarpers,Saito,NenciuReviewArticle},
\begin{equation}
h_{\subb}(\vt,\vtt) =\langle \delta_{\vt}, \op{H}_{\subb}\delta_{\vtt}\rangle= e^{ib\varphi(\vt,\vtt)} h_{0}(\vt,\vtt),
\label{defPeierlsPert}
\end{equation}
with
\begin{equation}  
\varphi(\vt,\vtt) = \frac{1}{2} (\gtt_1\gt_2-\gt_1\gtt_2)
=\frac{1}{2} \left[(\gtt_1,\gtt_2,0) \times (\gt_1,\gt_2,0)\right]_z, 
\label{PeierlsfaseFunktion}
\end{equation}
where $\left[\v{v}\right]_z$ denotes the
$z$-component of the $3$D vector $\v{v}$.
Note that $\varphi$ is anti-symmetric, that is $\varphi(\vt,\vtt)=-\varphi(\vtt,\vt)$. 
We denote by $\flux{\vt,\vt',\vt''}$ the magnetic flux of a field of
unit field-strength through the triangle generated by the sites $\vt$, $\vt'$ and $\vt''$. 
\[
\flux{\vt,\vt',\vt''} = \left[
 \frac{1}{2}\left[
( \vt'-\vt'' )\times(\vt-\vt')
\right]
 \right]_z.
\]
We note that the following identity holds:
\begin{equation}
 \flux{\vt,\vt',\vt''} =  \varphi(\vt,\vt') + \varphi(\vt',\vt'') +\varphi(\vt'',\vt). \label{fluxEquation}
\end{equation}

Finally, we define the position operators $\op{X}_{\nu}$, $\nu \in \{1,2\}$, densily defined on $\ell^2(\gT)$, by 
their action on the basis elements:
\begin{align*}
  \op{X}_{\nu} \delta_{\vt}  := \gt_{\nu} \delta_{\vt},\qquad \forall \vt=(\gt_1,\gt_2) \in \gT, \quad
  \nu \in \{1,2\}.
\end{align*}

\subsection{Derivation of the conductivity tensor from the Kubo formalism}
We use the same strategy as in \cite{R2} and \cite{FaradayRevisited}
in order to express the conductivity tensor as a local trace. 

We denote by $\mu$ the chemical potential and by $\beta =
\frac{1}{k_{\textup{B}} T}$ the inverse temperature. In the remote
past our system is at thermal equilibrium,
described by the Fermi-Dirac density operator:
\[
\opG{\varrho}_{0,\subN} =
f_{\subFD}(\op{H}_{\subb,\subN}), \quad f_{\subFD}(z) = \frac{1}{e^{\beta(z-\mu)}+1}.
\]

The system is perturbed by an incident monochromatic light-beam 
with a complex frequency $\omega$:
\begin{equation}\label{novemb1} 
 \omega =\omega_0 -\eta i,\quad \quad{\rm Re}(\omega)=\omega_0>0,\; {\rm Im}(\omega)=-\eta <0. \quad
\end{equation}
The electric electric field of the light-beam is parallel with the $x$-axis and given by
\[
\vec{E}(t) = (E_x(t), E_y(t)) =   \left( \Re \left(Ee^{i\omega
      t}\right), 0 \right)=Ee^{\eta t}\left( \cos({\omega_0
      t}), 0 \right), \quad \quad E \geq 0,\;t\leq 0,
\]
$\omega_0$ is supposed to satisfy condition \eqref{eq:omeganul}.
This electric field generates 
an external time-dependent potential term $\op{V}_{\subE}(t)$.
The full time-dependent Hamilton operator is now given by
\begin{equation} 
\op{H}_{\subE}(t)= \op{H}_{\subb} + \op{V}_{\subE}(t), \qquad
\op{V}_{\subE}(t)=\Re\left(Ee^{i\omega t} \right)\op{X}_1.
\end{equation}

To find the density operator at an arbitrary time $t \leq 0$, we need to solve the
dynamic (Liouville) equation with the initial condition at $t_0=-\infty:$ 
\begin{equation}
\label{rhoLiouville} 
i \frac{d{\opG{\varrho}}_{\subE,\subN}}{dt}(t)= [\op{H}_{\subE,\subN}\;, \opG{\varrho}_{\subE,\subN}(t)], \quad
\lim_{t\to-\infty}\opG{\varrho}_{\subE,\subN}(t)= f_{\textup{FD}}(\op{H}_{\subb,\subN}).
\end{equation}
Note that  at finite $N$ all operators are finite dimensional matrices.
The negative imaginary part of $\omega$ has the effect that
$\op{V}_{\subE}(\cdot)$ is absolutely integrable near $t = -\infty$.
Therefore equation~\eqref{rhoLiouville} has a unique solution which
can be iterated. Now we will identify the linear response coefficients.
\subsubsection{Current density}
The current operator in direction $\nu$ is defined as
\[
\op{j}_{\nu,\subb,\subN} := i
  [\op{H}_{\subE,\subN}\;, \op{X}_{\nu,\subN} ] = i[\op{H}_{\subb,\subN}\;, \op{X}_{\nu,\subN} ] \, \qquad
\nu = 1,2,
\]
which is independent of $E$ because $\op{V}_{\subE}(\cdot)$ commutes with $\op{X}_{\nu,\subN}$. 
The current density in the y-direction at $t=0$ is given by
\begin{align}
J_{2, \subb,  \subN}(E) &= \frac{1}{\abs{\gTN}}
\traceN{\opG{\rho}_{\subE,\subN}(t=0) \op{j}_{2,\subb, \subN}
}.\label{jto001}
\end{align}
It is well known that $J_{2, \subb, \subN}(E) $ has an analytic expansion in $E$ near zero,
and we will show that 
\begin{equation} 
J_{2,\subb, \subN}(E) = E\sigma_{21}(b,N) + \mathcal{O}(E^2). \label{eq:J2}
\end{equation}
The linear term in $E$ in formula~\eqref{eq:J2} defines the off-diagonal conduction element
that we seek to identify.
%
%
%
\section{Main results}\label{unu0}
\label{THM:MAIN}
The operator norm of $\op{H}_{\subb, \subN}$ is bounded from above by a constant uniform in $b$ and $N$, 
thus its spectrum lies in a sufficiently large closed interval $\mathbf{I}$, uniformly in $b$ and $N$.

\begin{figure}[h!tbp]
\begin{center}
\includegraphics[width=0.65\textwidth]{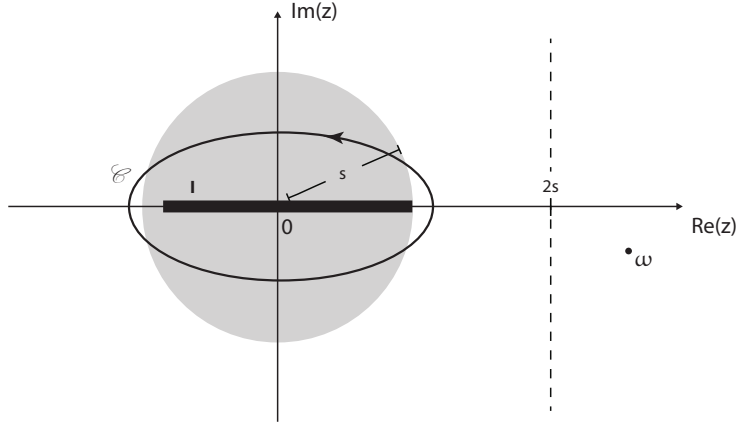}
\caption{\footnotesize{The imaginary part of $\omega$ should be so small that
$\abs{ \Im \omega } = \eta < \pi/\beta$. The path $\mathscr{C}$ should enclose
$\mathbf{I}$ and satisfy that $z \pm \omega \notin \mathbf{I}$. Such a $\mathscr{C}$ exists
if $\omega_0$ lies outside the interval defined in \eqref{eq:omeganul}, that is, if 
$\abs{\omega_0}$ is strictly larger than two times $s$, the spectral radius of $\op{H}_0$.}}
\label{fig:path}
\end{center}
\end{figure}

We define a smooth, simple and positively oriented closed path
$\mathscr{C}$ which encloses the above 
interval. Moreover all $z \in \mathscr{C}$ has to be close enough to the real line such 
that $f_{\subFD}(z)$ 
has no singularities inside $\mathscr{C}$, therefore suppose $\abs{ \Im \omega } = \eta < \pi/\beta$, see fig. \ref{fig:path}. 
A sufficient condition on $\abs{\Re \omega } = \abs{\omega_0}$, which will ensure existence of such a path $\mathscr{C}$, at least 
if $b$ is small enough, is
\begin{equation}
  \abs{\omega_0} > \{ 2 \abs{\xi} \; : \; \xi \in \sigma(\op{H}_{0})\},
   \label{eq:omeganul}
 \end{equation}
 that is, if  $\abs{\omega_0}$ is strictly larger than twice the spectral radius of $\op{H}_{0}$.
As a preliminary result, we shall see that $\sigma_{21}$ can be put in
the form:
\begin{align} \label{mainResult1}
&\sigma_{21}(b,N)=-\frac{\eta}{(\eta^2+\omega_0^2)\abs{\gTN}} 
\traceNtxt\left\{i[\op{j}_{2,\subb,\subN}, \op{X}_{1,\subN}
] f_{\subFD}(\op{H}_{\subN,\subb}) \right\}\\
&+ \Re 
\frac{1}{2\pi\omega\abs{\gTN}} 
\oint_{\mathscr{C}} \diffd z \;f_{\subFD}(z) 
\left(
\traceNtxt\left\{ 
\left(
  \op{H}_{\subN,\subb}  - z + \omega \right)^{-1} \op{j}_{1,\subb,\subN}  (\op{H}_{\subN,\subb}  -z)^{-1}
\op{j}_{2,\subb,\subN} \right\}+ [z \to z+\omega]\right),\nonumber
\end{align}
where $[z\rightarrow z+\omega]$ denotes a trace like the preceding one, just with $z$ substituted by $z+\omega$.
For $z \in \rho(\op{H}_b)$ define the operators on $\ell^2(\gT)$:
\begin{equation}
\op{D}_{\subb}(z) := \left( \op{H}_{\subb} -z + \omega \right)^{-1} \op{j}_{1,\subb} 
(\op{H}_{\subb}  -z)^{-1} \op{j}_{2,\subb}
\label{Doperator}
\end{equation}
where $\op{j}_{\nu,\subb} := i [\op{H}_{\subb} ,\op{X}_{\nu} ] , \, \nu = 1,2.$
We are now ready to state the main result of this paper.
 \begin{thm}
\label{thm:main} Assume that the real part of the frequency $\omega_0$
is large enough. Then the following statements hold true:

\noindent (i).
Assume that $\mathscr{C}_0$ contains the spectrum of $\op{H}_{\subb}$,
while $\mathscr{C}_0\pm\omega_0$ does not. Then the transverse
conductivity admits both the thermodynamic and the adiabatic limit, and we have:
\begin{equation}
\begin{split}
\label{mainResult2}
\sigma_{21}(b) &:=\lim_{\eta\to 0} \lim_{N\to \infty} \sigma_{21}(b,N) \\
&={\rm Re}\frac{1}{2\abs{\gUC}\pi\omega_0} 
\oint_{\mathscr{C}_0} \diffd z \;f_{\subFD}(z) 
\sum_{\gu \in \gUC} \left(\op{D}_{\subb}\kv{\vu,\vu;z} +
  \op{D}_{\subb}\kv{\vu,\vu;z+\omega_0}\right).
\end{split}
\end{equation}

\noindent (ii). In the standard nearest neighbour tight-binding model of graphene (see \eqref{decemb10}), $\sigma_{21}^G(0)=0$.

\noindent (iii).
The function $b \mapsto \sigma_{21}(b)$ is smooth and has an asymptotic
expansion in $b$ around $0$.  All the
derivatives of $\sigma_{21}$ at zero can be written only in terms of the fiber
operators associated to the Bloch decomposition of $\op{H}_0$. In particular, for the standard nearest neighbour tight-binding model of
graphene, all even Taylor coefficients are zero: 
 \begin{equation}
\diff[n]{\sigma_{21}^G}{b}(0)=0, \quad n=2p,\quad p \in \N.
\label{TaylorCoefficient}
\end{equation}
 \end{thm}

\vspace{0.5cm}

%
%
%
%
\section{Proof of Theorem  \ref{thm:main}(i)}\label{unu1}
%
%
\subsection{Derivation of formula \eqref{mainResult1}}
\label{sec:LargeNLimit}
If $\op{A}$ is a bounded operator, then we denote its expression in the interaction picture with
\begin{align*}
  \op{\tilde{A}} (t) = \exp\left(it\op{H}_{\subb,\subN}\right)\, \op{A}\,\exp\left(-it\op{H}_{\subb,\subN}\right).
\end{align*}
By standard perturbation theory, we can write 
\[
\opG{\varrho}_{\subN,\subE}(0)  =
f_{\subFD}(\op{H}_{\subN,\subb}) 
- i \int_{-\infty}^{0} \diffd s \left[ \tilde{\op{V}}_{\subN,\subE}(s)\, , \, 
f_{\subFD}(\op{H}_{\subN,\subb})
\right]  + \mathcal{O}(E^2)\,.
\]
Thus in the linear response approximation we set
\begin{equation} 
\opG{\varrho}_{\subN,\subE,\textup{lin}}(0)  =
f_{\subFD}(\op{H}_{\subN,\subb}) 
- i E \int_{-\infty}^{0} \diffd s\, \Re\left(e^{i\omega s}\right) \left[ \tilde{\op{X}}_{1\subN,\subE}(s)\, , \, 
f_{\subFD}(\op{H}_{\subN,\subb})
\right]\,. \label{rhonul}
\end{equation}
The value of the current density in the y-direction in the linear response regime is given by
 \[
 J_{2,\subb,\subN,\textup{lin}}(E,t=0) = \traceNtxt \left\{ \left( 
 f_{\subFD}(\op{H}_{\subN,\subb}) 
- i E \int_{-\infty}^{0} \diffd s\, \Re\left( e^{i\omega s}\right) \left[ \tilde{\op{X}}_{1\subN,\subE}(s)\, , \, 
f_{\subFD}(\op{H}_{\subN,\subb})
\right]\right)
 \frac{    \op{j}_{2,\subb,\subN}   }{  \abs{\gTN}  }
\right\}\,.
 \]
Using the
trace-cyclicity rule $\traceNtxt\{ [\op{A},\op{B}]\op{C} \} =
\traceNtxt\{ \op{B}[\op{C},\op{A}] \}$, the equilibrium current
$J_2(0)$ is shown to equal zero: 
\begin{equation*}
J_{2,\subb,\subN}(0) = \trace{[f_{\subFD}(\op{H}_{\subN,\subb}), \op{H}_{\subN,\subb}] \op{X}_{2,\subN}}
= 0.
\end{equation*}
By examining the formulae 
\eqref{jto001}, \eqref{eq:J2} and \eqref{rhonul}
we can single out the transverse conductivity term:
\begin{equation}
\opG{\sigma}_{21,\subN}(t=0) = 
-\frac{1}{\abs{\gTN}} \int_{-\infty}^{0} \diffd s \, 
\Re\left(e^{i\omega s} \right)\, \traceN{
i\left[ \tilde{\op{X}}_{1,\subb,\subN}(s)\, , \,
  f_{\subFD}(\op{H}_{\subN,\subb})\right]
\op{j_{2,\subb,\subN}} \label{eq:conductivityTensor1}
}. 
\end{equation}
The trace in the integrand of equation~\eqref{eq:conductivityTensor1}
is a real number, thus formula 
\eqref{eq:conductivityTensor1} can be re-expressed as
\begin{equation}
\opG{\sigma}_{21,\subN}(t=0)   =
-\frac{1}{\abs{\gTN}}\Re   \int_{-\infty}^{0} \diffd s\,e^{i\omega s}\,
\traceNtxt\left\{\, i [\op{j}_{2,\subb,\subN} \,,\, \UnulN{}{s} \op{X}_{1,\subN}
 \UnulN{-}{s}]\,  f_{\subFD}(\op{H}_{\subN,0}) \right\}.  \label{eq:condEq0012} 
\end{equation}
By partial integration  we re-express \eqref{eq:condEq0012} as
\begin{align}
\opG{\sigma}_{21,\subN}(t=0)   &=
-\frac{\eta}{(\eta^2+\omega_0^2)\abs{\gTN}} \traceNtxt\left\{i[\op{j}_{2,\subb,\subN}, \op{X}_{1,\subN}
] f_{\subFD}(\op{H}_{\subN,\subb}) \right\} \label{eq:condEq0013}
\\
&\quad + \Re \frac{1}{\abs{\gTN}}
\int_{-\infty}^{0} \diffd s\,
\frac{1}{i \omega} e^{i \omega s}
\diff{}{s}\left(\traceNtxt\left\{i[\op{j}_{2,\subb,\subN}, \UnulN{}{s} \op{X}_{1,\subN}
\UnulN{-}{s}] f_{\subFD}(\op{H}_{\subN,\subb}) \right\} \right). \nonumber
\end{align}
We note that
\begin{equation*} 
\diff{}{t} \left( 
\UnulN{}{t} \op{X}_{1,\subN}
\UnulN{-}{t}
\right) = 
\UnulN{}{t} \op{j}_{1,\subb,\subN}\UnulN{-}{t},
\end{equation*}
by the definition of $\op{j}_{1,\subb}$.
By using the trace-cyclicity and by noticing that
$f_{\subFD}(\op{H}_{\subN,\subb})$ 
and $e^{is(\omega + \op{H}_{\subN,\subb})}$ commute, we have
\[
\begin{split}
\opG{\sigma}_{21,\subN}(t=0) =& -\frac{\eta}{(\eta^2+\omega_0^2)\abs{\gTN}} \traceNtxt\left\{i[\op{j}_{2,\subb,\subN}, \op{X}_{1,\subN}
] f_{\subFD}(\op{H}_{\subN,\subb}) \right\}\nonumber \\
&+
\frac{1}{\abs{\gTN}} 
\Re\int_{-\infty}^{0} \left(
\frac{1}{\omega} 
\traceNtxt\left\{
e^{is \op{H}_{\subN,\subb})}\op{j}_{1,\subb,\subN} e^{-is(\op{H}_{\subN,\subb}- \omega)}
f_{\subFD}(\op{H}_{\subN,\subb}) \op{j}_{2,\subb,\subN} 
\right\}   \right.\\
&\qquad\qquad \qquad\qquad  \left.- \traceNtxt\left\{
e^{is(\omega + \op{H}_{\subN,\subb})}
f_{\subFD}(\op{H}_{\subN,\subb})  \op{j}_{1,\subb,\subN} e^{is \op{H}_{\subN,\subb}}
\op{j}_{2,\subb,\subN}) 
\right\}\right) \diffd s.
\end{split}
\]
We can use the Cauchy integral formula to express the operator $e^{is(\omega + \op{H}_{\subN,\subb})}
f_{\subFD}(\op{H}_{\subN,\subb})$ by a curve integral in the
complex plane, involving the
resolvent of $\op{H}_{\subN,\subb}$:
\[
e^{is(\omega +\op{H}_{\subN,\subb})}
f_{\subFD}(\op{H}_{\subN,\subb})
= \frac{i}{2\pi} \oint_{\mathscr{C}} e^{is(z+\omega)} f_{\subFD}
(z) (\op{H}_{\subN,\subb} -z)^{-1} dz,
\]
where the path $\mathscr{C}$ encloses,  but has no points in
common with, the (real, bounded) spectrum,
$\sigma(\op{H}_{\subb,\subN})$.
As mentioned, it is possible, given 
$\beta$, $\eta$, and $\omega_0$ satisfying \eqref{eq:omeganul}, to choose such a curve
$\mathscr{C}$ such that $\omega$ lies \emph{outside} $\mathscr{C}$, and such that 
$
f_{\subFD}(z) = (e^{\beta(z-\mu)} +1)^{-1} 
$
has no singularities inside $\mathscr{C}$. See figure~\ref{fig:path}. This leads to
\begin{equation}
\label{eq:simpleModel0001}
\begin{split}
&\opG{\sigma}_{21,\subN}(t=0)=-\frac{\eta}{(\eta^2+\omega_0^2)\abs{\gTN}} \traceNtxt\left\{i[\op{j}_{2,\subb,\subN}, \op{X}_{1,\subN}
] f_{\subFD}(\op{H}_{\subN,\subb}) \right\} \\
&+ \Re
\int_{-\infty}^{0} \diffd s\, \frac{1}{\omega\abs{\gTN}}  \left(
\traceNtxt \left\{ 
e^{is \op{H}_{\subN,\subb}} \op{j}_{1,\subb,\subN} 
\underbrace{\left( \frac{i}{2\pi} \oint_{\mathscr{C}} \diffd z \;e^{-is(z-\omega)}
  f_{\subFD}(z) (\op{H}_{\subN,\subb}-z)^{-1}
\right)}_{=e^{-is(\op{H}_{\subN,\subb}-\omega)} f_{\subFD}(\op{H}_{\subN,\subb})}
\op{j}_{2,\subb,\subN} 
\right\}\right. \\
&\qquad \quad-  \left. \traceNtxt\left\{
\underbrace{ 
  \left( \frac{i}{2\pi} \oint_{\mathscr{C}} \diffd z \,
     e^{is(z+\omega)}f_{\subFD}(z) (\op{H}_{\subN,\subb} - z
     )^{-1} \right)
}_{=e^{ is( \op{H}_{\subN,\subb}+\omega)}f_{\subFD}(\op{H}_{\subN,\subb})} 
 \op{j}_{1,\subb,\subN}\, e^{-is \op{H}_{\subN,\subb}} \, \op{j}_{2,\subb,\subN}
\right\}
\right).
\end{split}
\end{equation}
The two integrals, $\oint_{\mathscr{C}} \diffd z \dotsi$ and
$\int_{-\infty}^{0} \diffd s \dotsi$ in equation \eqref{eq:simpleModel0001} are both absolutely convergent, therefore we
can exchange integration order (Fubini). Furthermore,
as
$e^{\pm is(z \pm \omega)}$ is nothing but a complex scalar, we can
freely place this factor in the operator product,
\begin{equation*}
\begin{split}
&\opG{\sigma}_{21,\subN}(t=0)=-\frac{\eta}{(\eta^2+\omega_0^2)\abs{\gTN}} \traceNtxt\left\{i[\op{j}_{2,\subb,\subN}, \op{X}_{1,\subN}
] f_{\subFD}(\op{H}_{\subN,\subb}) \right\}\\ 
&+
\Re
\oint_{\mathscr{C}} \diffd z \;
\int_{-\infty}^{0} \diffd s\frac{1}{\omega\abs{\gTN}}  \left(
\traceNtxt\left\{ 
e^{is (\op{H}_{\subN,\subb} - z + \omega )} \op{j}_{1,\subb,\subN} 
\frac{i}{2\pi} 
  f_{\subFD}(z) (\op{H}_{\subN,\subb} -z)^{-1}
\op{j}_{2,\subb,\subN} \right\}\right. \\
&\qquad \quad- \left. \traceNtxt\left\{
 \frac{i}{2\pi} 
    f_{\subFD}(z) (\op{H}_{\subN,\subb} - z)^{-1}
\op{j}_{1,\subb,\subN} e^{is(z+\omega-\op{H}_{\subN,\subb} )}  \op{j}_{2,\subb,\subN}
\right\}
\right). 
\end{split}
\end{equation*}
Integrating with respect to $s$ we obtain (formula \eqref{mainResult1}):
\begin{align}
\begin{split}
\opG{\sigma}_{21,\subN}(t=0)&=-\frac{\eta}{(\eta^2+\omega_0^2)\abs{\gTN}} \traceNtxt\left\{i[\op{j}_{2,\subb,\subN}, \op{X}_{1,\subN}
] f_{\subFD}(\op{H}_{\subN,\subb}) \right\}\\
&+
\Re
\oint_{\mathscr{C}} \diffd z \frac{1}{2\pi\omega\abs{\gTN}} \;f_{\subFD}(z) 
\left(
\traceNtxt\left\{ 
\left(
  \op{H}_{\subN,\subb}  - z + \omega \right)^{-1} \op{j}_{1,\subb,\subN}  (\op{H}_{\subN,\subb}  -z)^{-1}
\op{j}_{2,\subb,\subN} \right\}\right. \\
&\qquad\qquad+ \left. \traceNtxt\left\{
(\op{H}_{\subN,\subb}  - z)^{-1}
\op{j}_{1,\subb,\subN}  \left(\op{H}_{\subN,\subb}  -z-\omega \right)^{-1}  \op{j}_{2,\subb,\subN} 
\right\}
\right).
\end{split}
\label{simpleModel051}
\end{align}
\subsection{Off-diagonal localization for resolvents}
In this subsection we only work with operators defined on
$\ell^2(\gT)$. It should be understood
that the results \ref{def:SH}-\ref{thm:CTprop} 
also hold true, even if $\gT$ is replaced with $\gTN$, uniformly in $N$.
\begin{definition}[Schur-Holmgren bound]
\label{def:SH}
For a linear operator $\op{A} \in B\left(\ell^2(\gT)\right)$ with a
kernel $a\kv{\v{x}, \v{x}'}$, we define the 
Schur-Holmgren bound by:
\begin{equation} 
\norm{\op{A}}_1 := \max \left\{ 
 \sup_{\v{x} \in \gT} \sum_{\v{x}' \in \gT} \abs{a\kv{\v{x}, \v{x}'}} , \sup_{\v{x}' \in \gT } \sum_{\v{x} \in \gT }  \abs{a\kv{\v{x}, \v{x}'}}
\right\} \label{eq:SchurHolmgreenSums}
\end{equation} 
If an operator has $\norm{\op{A}}_1 < \infty$ it is said to be \emph{Schur-Holmgren bounded}.
\end{definition}
The following bound is well known, and we give it without proof:
\begin{lemma}
\label{lem:SH-ON}
Let $\op{A} \in B\left(\ell^2(\gT)\right)$.  If $\norm{\op{A}}_1 < \infty$, then $\norm{\op{A}} \leq
\norm{\op{A}}_1 $, where $\norm{\op{A}}$ is the usual operator norm. 
\end{lemma}
\begin{definition}[Exponentially Almost Diagonal Operator]
\label{def:EADO}
Let $\op{A} \in B\left(\ell^2(\gT)\right)$. We say that 
$\op{A}$ is \emph{exponentially almost
  diagonal}, if there exist two constants $C_1,C_2$, both strictly positive, so that the kernel of $\op{A}$ satisfies 
\begin{equation} 
\abs{a\kv{\vt,\vtt}} \leq C_1 e^{-C_2\norm{\vt-\vtt}} \label{def:expLoc}
\end{equation}
for all $\vt,\vtt$ in $\gT$.
\end{definition}
The proof of the next lemma is straightforward and thus omitted.
\begin{lemma}
\label{thm:EXP-SH}
An exponentially almost diagonal operator is Schur-Holmgren bounded (and by lemma \ref{lem:SH-ON} bounded). 
\end{lemma}
We now show a property for exponentially almost diagonal operators,
which is a much simpler version of the Combes-Thomas estimate for
resolvents of continuous Schr\"odinger operators \cite{CT, RS4}.
%
%
%
\begin{proposition}[CT-property]
\label{thm:CTprop}
Let $z \in \rho(\op{A})$ where $\op{A} \in B\left(\ell^2(\gT)\right)$ is a
self-adjoint exponentially almost diagonal operator. Let constants $C_1$ and $C_2$ be defined as in
definition~\eqref{def:expLoc}.

Then the resolvent $(\op{A}-z)^{-1}$ is also exponentially almost
diagonal. That is, there exists two positive constants $C_3$
and $C_4$ such that the kernel of $(\op{A}-z)^{-1}$ fulfils
\begin{equation*} 
\abs{(\op{A}-z)^{-1} \kv{\vt,\vtt}} \leq C_3 e^{-C_4\norm{\vt-\vtt}}.
\end{equation*}
Consider the situation where $z$ is restricted to a closed subset $\mathscr{C} \subset
\rho(\op{A})$. Then $C_3$ and $C_4$ can be chosen uniformly in $\textup{dist}(\mathscr{C},\sigma(\op{A})\,)$.
\begin{proof}
  We only sketch the main ideas, see \cite{Nenciu3} for a related result. 
For $\alpha > 0$, and a fixed lattice point $\vt_{0}\in \gT$, define the operator
$\op{A}_{\alpha}:\ell^2(\gT) \to \ell^2(\gT)$ by
\begin{equation*} 
\op{A}_{\alpha} := e^{\alpha \norm{\cdot \,- \vt_{0}}} \op{A} 
e^{-\alpha \norm{\cdot    \,- \vt_{0}}}. 
\end{equation*}
It can be shown that
\begin{equation}
\norm{\op{A}-\op{A}_{\alpha}}_1
\,\autorightarrow{$\alpha\to 0$ }{}\, 0. \label{opInHlb0002}
\end{equation}
By the identity
\[
(\op{A}_{\alpha}-z) = (\identity -(\op{A}-\op{A}_{\alpha})(\op{A}-z)^{-1})(\op{A}-z),
\]
the operator
$(\op{A}_{\alpha}-z) $ is invertible for 
$\alpha$ small enough depending on $z$. We can write the equality
\[
e^{-\alpha \norm{\cdot\,-  \vt_{0}}} (\op{A}_{\alpha}-z) = (\op{A}-z) e^{-\alpha \norm{\cdot\,-  \vt_{0}}}, 
\]
which implies that
\[
(\op{A}-z)^{-1} e^{-\alpha \norm{\cdot\,-  \vt_{0}}} = e^{-\alpha \norm{\cdot\,-  \vt_{0}}} (\op{A}_{\alpha}-z)^{-1},
\]
and which gives
\[
e^{\alpha \norm{\cdot\,-  \vt_{0}}}  (\op{A}-z)^{-1} e^{-\alpha
  \norm{\cdot\,-  \vt_{0}}} = (\op{A}_{\alpha}-z)^{-1}.
\]
If we apply both sides on the basis element $\delta_{\vt_0}$ and then
take the scalar product with some $\delta_{\vt}$ we get:
\[
\langle \delta_{\vt}, e^{\alpha \norm{\cdot\,-  \vt_{0}}}  (\op{A}-z)^{-1} e^{-\alpha
  \norm{\cdot\,-  \vt_{0}}}\delta_{\vt_0}\rangle  =e^{\alpha
  \norm{\vt-  \vt_{0}}}(\op{A}-z)^{-1}(\vt,\vt_{0})= \langle
\delta_{\vt},(\op{A}_{\alpha}-z)^{-1}\delta_{\vt_0}\rangle,
\]
which means that 
$$e^{\alpha \norm{\vt-  \vt_{0}}}\left
  |(\op{A}-z)^{-1}(\vt,\vt_{0})\right |\leq \left \Vert
  (\op{A}_{\alpha}-z)^{-1}\right \Vert .$$
This formula can be used in order to get the desired estimate for the integral kernel. We do not give further details.
\end{proof}
\end{proposition}

%
%
%
%

\vspace{0.5cm}

The last preparatory results are concerned with the magnetic translations.
We define the magnetic translation operator as the operator that
transforms $\v{\psi} \in \ell^2(\gT)$ according to the rule:
\[
\left(\op{\mathcal{T}}_{\subb,\gg} \v{\psi} \right)(\vt) := e^{ib\varphi(\vt,
  \vg)}\psi(\vt-\vg), \qquad \textup{for all }\vg \in \gG, \vt \in \gT.
\]
The magnetic translations have certain properties which we list here without proof.
The inverse of the magnetic translation operator obeys
\[
\op{\mathcal{T}}_{\subb,\gg}^{-1} = \op{\mathcal{T}}_{\subb,-\gg}.
\]
The Hamilton operator $\op{H}_{\subb}$ commutes with
$\op{\mathcal{T}}_{\subb,\gg}$ for all $\gg \in \gG$.
The same property is true for the current operator whose integral kernel is
\begin{equation*} 
  \op{j}_{\nu,\subb}\kv{\vt,\vtt} = i h_{\subb}\kv{\vt,\vtt}(\gtt_{\nu}-\gt_{\nu}) =
  e^{ib\varphi(\vt,\vtt)} \op{j}_{\nu,0}(\vt,\vtt), \quad \nu \in \{1,2\}.
\end{equation*}

\subsection{Proof of the thermodynamic and adiabatic limits}
Let us briefly discuss the first term appearing in
\eqref{mainResult1}, that is 

$$\frac{\eta}{(\eta^2+\omega_0^2)\abs{\gTN}} 
\traceNtxt\left\{i[\op{j}_{2,\subb,\subN}, \op{X}_{1,\subN}
] f_{\subFD}(\op{H}_{\subN,\subb}) \right\}.$$
We have the identity: 
$$i[\op{j}_{2,\subb,\subN}, \op{X}_{1,\subN}
](\vt,\vtt)= i \chN(\vt) h_b(\vt,\vtt) (y_2-x_2)(y_1-x_1)\chN(\vtt),$$
which defines a bounded operator due to the localization properties of
$h_0$. This means in particular that $\traceNtxt\left\{i[\op{j}_{2,\subb,\subN}, \op{X}_{1,\subN}
] f_{\subFD}(\op{H}_{\subN,\subb}) \right\})/\abs{\gTN}$ is bounded in
$N$ uniformly in $\eta$, thus after the adiabatic limit this term will
disappear anyway. That is why we only treat in detail the second term
of \eqref{mainResult1}.

\begin{figure}[h!tbp]
\begin{center}
\includegraphics[width=0.5\textwidth]{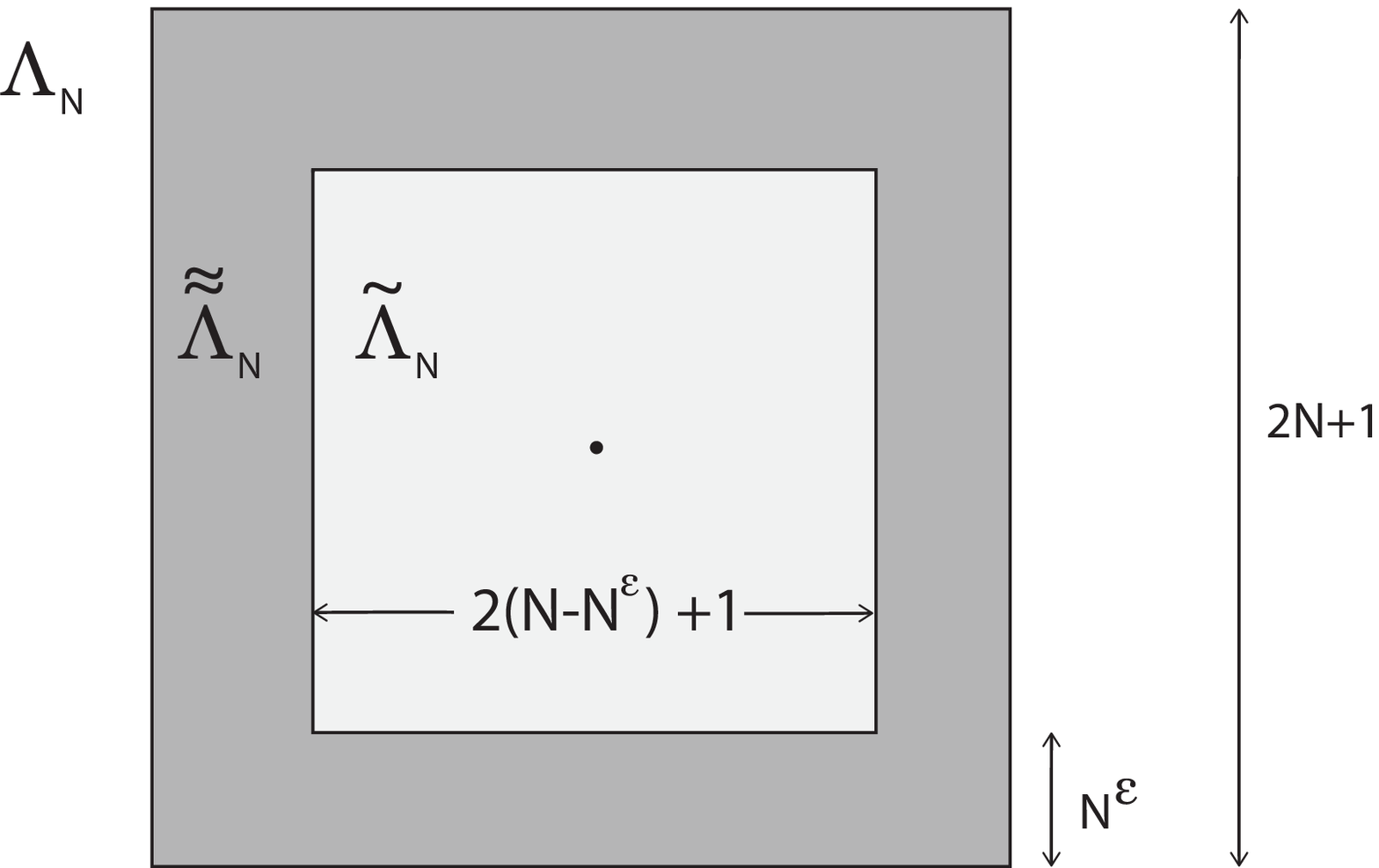}
\caption{\footnotesize{$\gTN$ is split into a core part, $\gTNt$,  $2(N-N^{\epsilon})+1$ unit cells
wide and an edge region $\gTNtt$ of width $N^{\epsilon}$ unit cells. $0<\epsilon<1$}}
\label{fig:termodynamic1}
\end{center}
\end{figure}

First, we need to introduce some notation. 
Let $0<\epsilon < 1$. We divide our finite box, $\gTN$, into an edge
region $\gTNtt$ of width $[N^{\epsilon}]$ unit cells, and a remaining
core part, $\gTNt$, see figure \ref{fig:termodynamic1}. We have that
$\gTNtt = \gTN \setminus \gTNt$. For practical reasons we work with $N^{\epsilon}$ instead of its integer part.
The number of unit cells in $\gTNtt$ is $N^{1+\epsilon}$, which means $\frac{\abs{\gTNtt}}{N^2} \to 0$ as $N$ tends to infinity. 
As a consequence, the number unit cells in $\gTNt$ behaves like $4N^2$ for large $N$.
\subsubsection{Geometric perturbation theory}
To simplify notation, we introduce a shorthand for the characteristic
functions  
$\chNt = \chi_{\gTNt}$, $\chNtt~=~\chi_{\gTNtt}$.
We now introduce an auxilliary operator by:
\[
\op{A}_{\subb, \subN}(z) = 
\chNt\left( \op{H}_{\subb} - z\right)^{-1} \chNt
+ \chNtt \left( \op{H}_{\subb,\subN} - z\right)^{-1} \chNtt, \quad z \in \mathscr{C}.
\]
If we multiply $\op{A}_{\subb, \subN}(z)$ on the left by $\left( \op{H}_{\subb, \subN}  - z\right)$, we have
\begin{align} 
\left( \op{H}_{\subb, \subN}  - z\right)\op{A}_{\subb,\subN}(z) &=
\left( \op{H}_{\subb, \subN}  - z\right)\chNt \left( \op{H}_{\subb}  - z\right)^{-1} \chNt
+
\left( \op{H}_{\subb, \subN}  - z\right)\chNtt \left( \op{H}_{\subb, \subN}  - z\right)^{-1} \chNtt. \label{termo001}
\end{align}
Note that for large enough $N$, the distance between $\gTNt$ and $\gT \setminus \gTN$ becomes larger than the interaction range of $\op{H}_{\subb,\subN}$. This
implies that $\left( \op{H}_{\subb, \subN}  - z\right)\chNt = \left( \op{H}_{\subb}  - z\right)\chNt$. Therefore, if we use this in  the first term of formula~\eqref{termo001}, and using that $\chNt + \chNtt = \chN$, \eqref{termo001} becomes:
\begin{align*} 
\left( \op{H}_{\subb, \subN}  - z\right)\op{A}_{\subb,\subN}(z)  &=
\chN + [\op{H}_{\subb}, \chNt]  \left( \op{H}_{\subb}  - z\right)^{-1} \chNt
+ [\op{H}_{\subb,\subN}, \chNtt]  \left( \op{H}_{\subb, \subN}  - z\right)^{-1} \chNtt,
\end{align*}
which is equivalent with:
\begin{align} 
\left( \op{H}_{\subb,\subN} - z \right)^{-1}    &= \op{A}_{\subb,\subN}(z)  - \left( \op{H}_{\subb,\subN} - z \right)^{-1} \op{B}_{\subb,\subN}(z),\label{termo004}
\end{align}
where:
\[
\op{B}_{\subb,\subN}(z) = [\op{H}_{\subb}, \chNt]  \left( \op{H}_{\subb}  - z\right)^{-1} \chNt
+ [\op{H}_{\subb,\subN}, \chNtt]  \left( \op{H}_{\subb, \subN}  - z\right)^{-1} \chNtt.
\]
We insert \eqref{termo004} in formula~\eqref{simpleModel051} obtaining
several terms. 
We claim that only the following term contributes in the large $N$ limit:
\begin{equation}
\Re\frac{1}{2\pi\omega\abs{\gTN}} 
\oint_{\mathscr{C}} \diffd z \;f_{\subFD}(z) 
\left(
\traceNtxt\left\{
\chNt \left( \op{H}_{\subb} -z + \omega \right)^{-1} \chNt\op{j}_{1,\subN} 
\chNt(\op{H}_{\subb}  -z)^{-1} \chNt
\op{j}_{2,\subN} \right\}
+ [z\rightarrow z+\omega]
\right).
\label{termo022}
\end{equation}
The other terms in the expansion of formula~\eqref{simpleModel051} have factors of type
 $\chNtt\left(\op{H}_{\subb,\subN} - z \right)^{-1}$,
 $[\op{H}_{\subb}, \chNt]$ or $[\op{H}_{\subb,\subN}, \chNtt]$. In the
 large $N$ limit terms having these factors vanish, which we explain in the following.

To begin with, let us choose one such term which up to trace cyclicity can be written as
\[
\frac{1}{\abs{\gTN}}\traceNtxt\left\{ 
[\op{H}_{\subb}, \chNt] \op{Q}_{b, \subN}(z) \right\}, \quad z \in \mathscr{C},
\]
where $\op{Q}_{b, \subN}(z)$ is bounded uniformly in $b$, $N$, and $z$. Because the operator $\op{H}_{\subb}$ is short-range, we can find
a projection $\op{P}_{\subN}$ whose corresponding subspace has a dimension $D\sim N^{1+\epsilon}$, such that $\op{P}_{\subN}[\op{H}_{\subb}, \chNt]=[\op{H}_{\subb}, \chNt]$. We then use the inequality 
\[
\abs{\traceNtxt\left\{[\op{H}_{\subb}, \chNt] \op{Q}_{b, \subN}(z) \right\}}
\leq
C \traceNtxt\left\{ \op{P}_{\subN} \right\} \sim N^{1+\epsilon},
\]
where $C$ is a constant uniform in $b$, $N$, and $z$. Thus by dividing by $N^2$, it will converge to zero. The same type of proof applies for all other terms containing 
 $\chNtt$ which can already play the role of $\op{P}_{\subN}$.

We now consider the term given by formula~\eqref{termo022}, and we
want to show that we can 
replace $ \chNt\op{j}_{1,\subN} 
\chNt$ with $\op{j}_{1}$. If we write 
\[
\chNt = \identity -(\identity - \chNt),
\]
then the trace in \eqref{termo022} can be written as:
\begin{equation}
\traceNtxt\left\{
\chNt \left( \op{H}_{\subb} -z + \omega \right)^{-1} \left(\identity -(\identity - \chNt)\right) \op{j}_{1} 
\left(\identity -(\identity - \chNt)\right) (\op{H}_{\subb}  -z)^{-1} 
\chNt \op{j}_{2} \chNt  \right\}
+ [z\rightarrow z+\omega].
\label{termo202}
\end{equation}
This trace, and thereby formula \eqref{termo022}, can now be expanded into several terms, all but one containing at least one factor of the type
$(\identity - \chNt)$. The terms containing $(\identity - \chNt) = (\identity - \chN + \chNtt)$ as a factor can be written (up to trace cyclicity) in the form
\begin{equation}
\Re\frac{1}{2\pi\omega\abs{\gTN}} 
\oint_{\mathscr{C}} \diffd z \;f_{\subFD}(z) 
\traceNtxt\left\{
(\identity - \chN)
\op{E}_{\subb}(z)  \chNt  \op{Q}_{b, \subN}^{(1)}(z)+ \chNtt\op{Q}_{b, \subN}^{(2)}(z) 
\right\}, \label{termo201}
\end{equation}
where $\op{E}_{\subb}(z)$ is an exponentially almost diagonal operator and $\op{Q}^{(1)}_{b, \subN}(z)$ and $\op{Q}^{(2)}_{b, \subN}(z)$ are bounded, uniformly in $b$, $N$ and $z$. The term $\chNtt\op{Q}^{(2)}_{b, \subN}(z)$ vanishes as $N$ tends to infinity by previous arguments.
Now consider the term containing $(\identity - \chN)\op{E}_{\subb}(z)\chNt$. 
The kernel $k(\vt,\vtt)$ of this operator is zero unless $\norm{\vt-\vtt}>N^{\epsilon}$, see figure~\ref{fig:termodynamic1}. One can easily show that the trace-norm of this operator 
is bounded from above by $C_1e^{-C_2N^{\epsilon}}$, for some positive constants $C_1$ and $C_2$. Thus this term will not give a contribution to the 
thermodynamic limit.
The only remaining contribution from formula \eqref{termo022} is
\begin{align}
\begin{split}
&\Re\frac{1}{2\pi\omega\abs{\gTN}} 
\oint_{\mathscr{C}} \diffd z \;f_{\subFD}(z) 
\left(
\traceTxt\left\{
\chNt \left( \op{H}_{\subb} -z + \omega \right)^{-1} \op{j}_{1,\subb} 
(\op{H}_{\subb}  -z)^{-1} 
\op{j}_{2,\subb}\chNt \right\}\right.
\\
&\qquad\qquad+ \left. \traceTxt\left\{
\chNt 
(\op{H}_{\subb}  - z)^{-1} \op{j}_{1,\subb}
\left(\op{H}_{\subb}  -z-\omega \right)^{-1} \op{j}_{2,\subb} \chNt 
\right\}
\right).
\end{split}
\label{termo022b}
\end{align}
Using the operators defined in~\eqref{Doperator}, the previous formula can be re-written as
\begin{equation}
\Re\frac{1}{2\pi\omega\abs{\gTN}} 
\oint_{\mathscr{C}} \diffd z \;f_{\subFD}(z) 
\left(
\sum_{\vt \in \gTNt}\op{D}_{\subb}\kv{\vt,\vt;z}
+ 
\op{D}_{\subb}\kv{\vt,\vt;z+\omega}\right).
\label{termo040}
\end{equation}
$\op{D}_{\subb}(z)$ is a product of operators 
which commute with magnetic translations.
This implies that the diagonal elements of its integral kernel define a periodic function, that is
for any $\vu \in \gUC$ and $\vg \in \gG$, we have that
\[
\op{D}_{\subb,\pm}(\vu+\vg,\vu+\vg ; z) = 
\op{D}_{\subb,\pm}(\vu,\vu;z) .
\]
Now the proof of \eqref{mainResult2} is straightforward and the
thermodynamic and adiabatic limits are proved.
\section{Proof of Theorem \ref{thm:main}(ii)}\label{unu2}

Here we need to compute the transverse conductivity component at
$b=0$ and prove that it gives zero if we work with the operator \eqref{decemb10}. We will do this computation for a
general nearest neighbor model, and use the graphene model only at the end.

\subsection{The Bloch-Floquet representation}
In order to fix notation, 
we define $\hlbG_{\subF}:=\int_{\gBZ}^{\oplus}\diffd^2k\,\, \ell^2(\gUC) $.
The Floquet unitary~\cite{RS4} $\op{U}:\ell^2(\gT) \to \hlbG_{\subF}$
takes vectors $\v{\psi}$ from $\ell^2(\gT)$ into
$\hlbG_{\subF}$,
and is given by the well-known formula
\begin{equation}
(\op{U}\v{\psi})(\v{k},\vu) = \frac{1}{\sqrt{\abs{\gBZ}}}\sum_{\gg \in \gG} \exp(-i\v{k}\cdot\vg) \psi(\vu + \vg),
\quad \v{k} \in \gBZ, \vu \in \gUC. \label{floquetTransform001}
\end{equation}
If $\op{A}$ is any bounded self-adjoint operator commuting with the
translations induced by the Bravais lattice $\Gamma$, we have that 
$\op{U}\op{A}\op{U}^*:\hlbG_{\subF}\to \hlbG_{\subF}$ is given by
\[
\op{U}\op{A}\op{U}^* = \int_{\gBZ}^{\oplus} a_{\subF}(\v{k}) \,\diffd \v{k},
\]
where the fibers have the kernels (we assume that $a\kv{\vu+\vg,\vut}$
has a sufficiently fast decay with $\vg$):
\begin{equation} 
a_{\subF}(\vu,\vut;\v{k}) = \sum_{\gg \in \gG} e^{-i\v{k}\cdot\vg}
a\kv{\vu+\vg,\vut}. \label{k-spaceCrystalHamilton}
\end{equation}
We denoted the number of sites of $\Omega$ with
$\nu_\Omega$. For each $\v{k}\in \gBZ$,
$a_{\subF}(\v{k})$ is self-adjoint and has  
$\nu_\Omega$ real eigenvalues. 
Each matrix-component of $a_{\subF}(\v{k})$, viewed as a function of $\v{k}$, extends 
to a $\gG^*$-periodic $C^{\infty}(\R^2)$ function. 
We order these eigenvalues in increasing order:
\[
\epsilon_1(\v{k}) \leq \epsilon_2(\v{k}) \leq \dots \leq \epsilon_{\nu_\Omega}(\v{k}).
\]

In order to have periodic boundary conditions in the fibers, we modify
\eqref{floquetTransform001} with a 
complex phase and define:
\begin{equation}
\op{U}_{\subF} : \ell^2(\gT) \to \hlbG_{\subF}, \qquad
\left( \op{U}_{\subF}  \v{\psi}\right)(\v{k}, \vu) = \sum_{\gg \in \gG} e^{-i\v{k}\cdot(\vu+\vg)}\psi(\vu+\vg), ~\label{Uf}
\end{equation}
for all $\v{\psi} \in \ell^2_c(\gT)$.
Accordingly $\op{U}_{\subF} \op{H}_0 \op{U}^*_{\subF} : \hlbG_{\subF} \to \hlbG_{\subF}$ 
has the fibers 
\[
\tilde{h}_0\kv{\gu,\gut;\v{k}} = \sum_{\vg \in
  \gG} h_0\kv{\gu+\vg,\gut} e^{-i\v{k}\cdot(\gu+\vg-\gut)}.
\]

If we differentiate the fiber with respect to the first component
of $\v{k}$, we have
\begin{equation}
\pdiff{}{k_1} \tilde{h}_0\kv{\vu,\vut;\v{k}} 
= -\sum_{\gg \in \gG} e^{-i \v{k}\cdot ( \vu + \vg - \vut )}\; i (
\gu_1 + \gg_1-\gut_1  ) h\kv{\vu + \vg, \vut} \label{mik1001}.
\end{equation}
The expression \eqref{mik1001} is nothing but the fiber of the
transformed current operator $\op{U}_{\subF}\,( i[\op{H},\op{X}_1])\,\op{U}^*_{\subF}$. 

\begin{figure}[h!tbp]
\begin{center}
\includegraphics[width=0.8\textwidth]{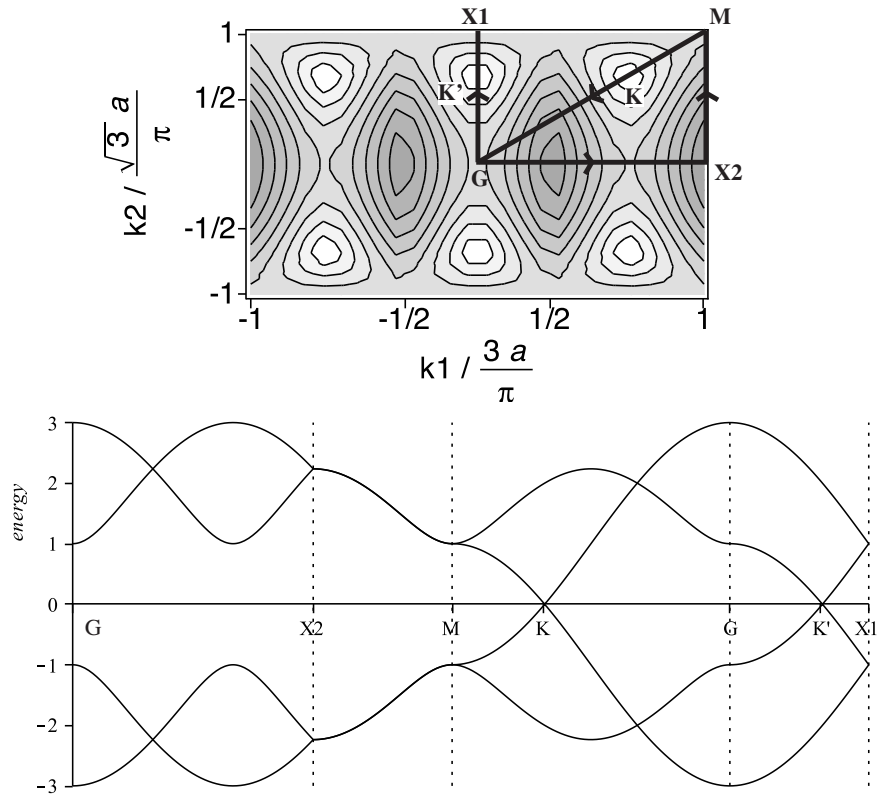}
\caption{\footnotesize{The bandstructure of the standard nearest neighbour tight-binding model of
    Graphene. Top: a 
figure of the first Brillouin zone, with the level-curves of
$e_3(k_1,k_2)$ plotted, also showing some points of interest: $G$ (origin)
$X1,X2,M,K$ and $K'$. $K$ and $K'$ are the two ``Dirac points''.
Bottom: the plot of the Bloch energies 
along the path indicated in the top figure. }}
 \label{fig:bandstructure}
\end{center}
\end{figure}

In particular, for the graphene-Hamiltonian \eqref{decemb10} we have $\nu_\Omega=4$, with the numbering of the basis positions given in \eqref{basisVectors}:
\small
\begin{equation*}
  \tilde{h}_0^G(\v{k}) = 
  \left[ \begin {array}{cccc} 
  0 & e^{i k_1 a} & 0 & 2 e^{- \frac{i k_1 a}{2}} \cos \left( \frac{\sqrt {3} k_2 a}{2} \right) 
\\ 
\noalign{\medskip}{  e^{-i k_1 a}  } & 0  &   2 e^{ \frac{i k_1 a}{2}} \cos \left( \frac{\sqrt {3} k_2 a}{2} \right)  &0
\\ \noalign{\medskip} 0 &    2 e^{- \frac{i k_1 a}{2}} \cos \left( \frac{\sqrt {3} k_2 a}{2} \right)  &0&  e^{i k_1 a} 
\\ \noalign{\medskip}   2 e^{ \frac{i k_1 a}{2}} \cos \left( \frac{\sqrt {3} k_2 a}{2} \right)  & 0 &  e^{-i k_1 a} &0
\end {array} \right]. 
\end{equation*}
\normalsize
We see that all matrix elements are \emph{even} in $k_{2}$. The
bandstructure is given 
by the eigenvalues of $\tilde{h}_0^G(\v{k})$  which are shown in figure \ref{fig:bandstructure}.

Now if we have two operators $\op{A}, \op{B} \in B(\ell^2(\gT))$, both $\gG$-periodic,
then it holds that:
\begin{equation}
(\op{A}\op{B})\kv{\vu,\vut;\v{k}} = \sum_{\vu' \in \gUC}
a\kv{\vu,\vu';k} b \kv{\vu',\vut;k}.
\label{kProdRule}
\end{equation}
When $b$ is equal to zero, formula \eqref{mainResult2} becomes:
\begin{align*}
\sigma_{21}(0)&=\frac{1}{2\abs{\gUC}\pi\omega_0} \Re
\oint_{\mathscr{C}} \diffd z \;f_{\subFD}(z) 
\sum_{\gu \in \gUC} \left(\left[
(\op{H}_{0}-z+\omega_0)^{-1} \op{j}_{1,0}
(\op{H}_{0}-z)^{-1}\op{j}_{2,0} 
\right]  \kv{\vu,\vu} + [z \rightarrow z + \omega_0]\right). \label{konstantConduction}
\end{align*}
In order to shorter notation, we denote by $r_{0}\kv{z,\v{k}}=(\tilde{h}_0\kv{\v{k}}-z)^{-1}$
the fiber of the resolvent $(\op{H}_{0}-z)^{-1}$. 
Using \eqref{kProdRule} we can write  
\begin{equation}
\begin{split}
&\sigma_{21}(0)=\frac{1}{2\abs{\gUC}\pi\omega_0} \sum_{\gu,\gu',\gu'',\gu''' \in \gUC} \frac{1}{\abs{\gBZ}}
\int_{\gBZ}\diffd^2k\quad \Re
\oint_{\mathscr{C}} \diffd z \;f_{\subFD}(z) 
\\
&\qquad\left(
r_{0}\kv{\vu,\vu';z-\omega_0,\v{k}} 
\pdiff{}{k_1}\tilde{h}_0\kv{\vu',\vu'';\v{k}}
r_{0}\kv{\vu'',\vu''';z,\v{k}} 
\pdiff{}{k_2}\tilde{h}_0\kv{\vu''',\vu;\v{k}}
+  \right.\\
&\left.\qquad +
r_{0}\kv{\vu,\vu';z,\v{k}} 
\pdiff{}{k_1}\tilde{h}_0\kv{\vu',\vu'';\v{k}}
r_{0}\kv{\vu'',\vu''';z+\omega_0,\v{k}} 
\pdiff{}{k_2}\tilde{h}_0\kv{\vu''',\vu;\v{k}}
\right).
\end{split} \label{nulTerm}
\end{equation}
Specializing this formula for $\tilde{h}_0^G\kv{\v{k}}$, we see that by differentiating with respect to $k_2$ and then
integrating with respect to $z$, the total integrand for the $\v{k}$ integral becomes an odd function of $k_2$. 
When we integrate $k_2$ on the symmetric Brillouin zone, the integral
giving $\sigma_{21}^G(0)$ equals zero.

%
%
%

\section{Proof of Theorem \ref{thm:main} {\rm (iii)}}\label{unu3}
This section is where the gauge-invariant magnetic perturbation 
theory \cite{NenGI,MPT1,MPT2,MPT3}
plays a crucial role. The main idea behind 
this method is to express the resolvent of $H_b$ as a norm convergent series in $b$: 
\begin{equation*}
  (H_{b} -z)^{-1} = \sum_{j=0}^{\infty} b^j T_{j,b}(z), 
  \quad 
  z \in \rho(H_{b}),
\end{equation*}
where the coefficient operators $ T_{j,b}(z)$ still depend on the magnetic field, but
only through unimodular exponential factors. For an introduction 
to gauge invariant magnetic perturbation theory, see \cite{NenGI}.

For any $z \in \rho(\op{H}_0)$, we define the operator
 $\op{S}_{\subb}(z)$ 
by its kernel
\begin{equation}
s_{\subb}(\vt,\vtt;z) = e^{ib\varphi(\vt,\vtt)}\left((\op{H}_{0} - z)^{-1}\right)(\vt,\vtt)
.\label{sDef}
\end{equation}
Notice that a Schur-Holmgren estimate shows that when $z$ is
restricted to a compact set $\mathscr{C} \subset
\rho(\op{H}_{0})$, then  
$\norm{\op{S}_{\subb}(z)} \leq C_{0}(\mathscr{C})$ for some positive
constant 
$C_{0}(\mathscr{C})$, uniformly in $z \in \mathscr{C}$. 
By denoting with $\identity$ the identity operator, we define
\begin{equation}
\op{K}_b(z)=(\op{H}_{\subb} - z)\op{S}_{\subb}(z) - \identity, \label{SgangeH001}
\end{equation}
where the operator $\op{K}_b(z)$ has the integral kernel:
\[
k_{\subb}\kv{\vt,\vt';z} = e^{ib\varphi(\vt,\vt')} \sum_{\vtt}\left( e^{ib\,\flux{\vt,\vtt,\vt'}} -1 \right)
h_0\kv{\vt,\vtt}s_0\kv{\vtt,\vt';z}.
\]
Using the exponential localization of the above integral kernels,
together with the estimate 
$$|\flux{\vt,\vtt,\vt'}|\leq ||\vt-\vtt||\;||\vtt-\vt'||, $$
then a Schur-Holmgren estimate applied to $\op{K}_{\subb}(z)$ shows that 
\begin{equation}
\sup_{z\in \mathscr{C}}\left\{\norm{\op{K}_{\subb}(z)} \right\}\leq b\, C(\mathscr{C}), \label{KbNormBound2}
\end{equation}
for some positive constant $C(\mathscr{C})$, uniformly in $z \in
\mathscr{C}$. The constant $C(\mathscr{C})$ only depends on the distance between
$\mathscr{C}$ and $\sigma(\op{H}_{0})$.
%
%

%
The next lemma is a direct consequence of the above estimates and
recovers a well-known result about the spectrum stability of
$\op{H}_{\subb}$. We state it here without other details; see \cite{Bell, HoriaHarpers} 
for much stronger results.  
\begin{lemma}
\label{thmStability01}
Let $\mathscr{C} \subset \rho(\op{H}_{0} )$ be any compact set. Then there exists
$b_{\mathscr{C}}>0$, sufficiently small, such that $\mathscr{C}~\subset~\rho(\op{H}_{\subb})$ for all $0~\leq~b~\leq~b_{\mathscr{C}}$.
\end{lemma}

\vspace{0.5cm}

From now on $\mathscr{C}$ is the integration contour in the formula
giving the conductivity, and $z\in \mathscr{C}$. If $b$ is small
enough we can write
\begin{align}
(\op{H}_{\subb} - z )^{-1}&=\op{S}_{\subb}(z) [\identity +
\op{K}_{\subb}(z)]^{-1}=\op{S}_{\subb}(z)-\op{S}_{\subb}(z) [\identity
+ \op{K}_{\subb}(z)]^{-1}\op{K}_{\subb}(z)\nonumber \\
&= \op{S}_{\subb}(z) - (\op{H}_{\subb}-z)^{-1} \op{K}_{\subb}(z).
\label{HbResolv}
\end{align}
We can iterate this and obtain
\begin{equation}
(\op{H}_{\subb}-z)^{-1} = \op{S}_{\subb}(z) -
\op{S}_{\subb}(z)\op{K}_{\subb}(z) + \op{\mathcal{R}}_{\subb}(z),
\label{resolventWithb}
\end{equation}
with the definition of the remainder term
\[
\op{\mathcal{R}}_{\subb}(z)  := (\op{H}_{\subb}-z)^{-1} \op{K}_{\subb}^{2}.
\]
Now both factors defining this remainder are exponentially localized,
and standard estimates lead to: 
\[
\sup_{z \in \mathscr{C}} \abs[\big]{\op{\mathcal{R}}_{\subb}(\vt,\vtt;z)} \leq
b^2 \, c_{\mathcal{R}}e^{-\tilde{c}_{\mathcal{R}} \norm{\vt-\vtt}}, \quad \vt, \vtt \in \gT,
\]
for some positive constants $\tilde{c}_{\mathcal{R}}$ and
$c_{\mathcal{R}}$. This shows that the remainder is also 
exponentially almost diagonal, thus there exists a constant $C_{\mathcal{R}}(\mathscr{C})>0$ such that
\begin{equation}
\norm{\op{\mathcal{R}}_{\subb}(z)} \leq b^{2} C_{\mathcal{R}}(\mathscr{C}).
\label{RemainderNorm}
\end{equation}

\subsection {The first derivative of $\sigma_{21}(b)$}
Now we seek to identify the linear part in $b$
of~\eqref{mainResult2}. Consider again \eqref{mainResult2}: 
\begin{align}\label{bConduction}
&\sigma_{21}(b)\\
&=\Re\frac{1}{2\abs{\gUC}\pi\omega_0} 
\oint_{\mathscr{C}} \diffd z \;f_{\subFD}(z) 
\sum_{\gu \in \gUC} \left[
(\op{H}_{\subb}-z+\omega_0)^{-1} \op{j}_{1,\subb}
(\op{H}_{\subb}-z)^{-1}\op{j}_{2,\subb} 
+(z \rightarrow z+\omega_0)\,\right]  \kv{\vu,\vu}. \nonumber
\end{align}
Using formula~\eqref{resolventWithb}, we see that by substituting
$(\op{H}_{\subb}-z)^{-1}$ with 
$\op{S}_{\subb}(z) - \op{S}_{\subb}(z)\op{K}_{\subb}(z)$, the error we
make is of order $b^2$ and this remainder cannot contribute to the first order
derivative at $b=0$. Therefore:
\begin{align}
&b\sigma_{21}^{(1)}+\mathcal{O}(b^2)=\Re\frac{1}{2\abs{\gUC}\pi\omega_0} 
\oint_{\mathscr{C}} \diffd z \;f_{\subFD}(z) \label{bConduction2}\\
&\sum_{\gu \in \gUC} \left[
\left( \op{S}_{\subb}(z-\omega_0)-\op{S}_{\subb}(z-\omega_0)\op{K}_{\subb}(z-\omega_0)\right)
\op{j}_{1,\subb}
\left[ \op{S}_{\subb}(z) -\op{S}_{\subb}(z)\op{K}_{\subb}(z) \right]
\op{j}_{2,\subb} +(z \to z + \omega_0) 
\right]  \kv{\vu,\vu}. \nonumber
\end{align}
We will now sketch the calculation of the trace over the basis for
a given $z$.  The following computations hold uniformly
in $z \in \mathscr{C}$ and $0 \leq b \leq b_{\mathscr{C}}$. 
We introduce the following shorthands:
\begin{alignat*}{4}
\sz := \op{S}_{\subb}(z),  
&\quad\szm := \op{S}_{\subb}(z-\omega_0), 
\quad\qquad\qquad &\textup{"S-type"}\\ 
\sk := \op{S}_{\subb}(z)\op{K}_{\subb}(z), 
&\quad\skm := \op{S}_{\subb}(z-\omega_0)\op{K}_{\subb}(z-\omega_0), 
\qquad &\textup{"SK-type"}. 
\end{alignat*}
For $\vu \in \gUC$ consider the element 
\begin{equation}
\left[\;
\left(\szm - \skm \right)\op{j}_{1b} \left(\sz - \sk \right) \op{j}_{2b} +
(z\to z+\omega_0)
\;\right]\kv{\vu,\vu}. \label{HejsaA}
\end{equation}
Let us expand it:
\begin{equation}
  [\szm \op{j}_{1b} \sz\op{j}_{2b}  -\szm \op{j}_{1b} \sk\op{j}_{2b} - \skm\op{j}_{1b}\sz\op{j}_{2b} 
+ \skm\op{j}_{1b} \sk \op{j}_{2b}+(z \to z+\omega_0)]\kv{\vu,\vu}. 
\label{hejsa}
\end{equation} 

To show the method of calculation, we first consider the operator product of two of ``S-type'' operators.
\[
\left[\op{S}_{\subb}(z-\omega_0) \op{j}_{1,\subb} \op{S}_{\subb}(z) \op{j}_{2,\subb} \right]\kv{\vu,\vu}, \quad \vu \in \gUC.
\]
Here, the $b$-dependence of the integral kernel appears only through
the exponential phases. Denoting by
\[
\FL(\vu,\vt',\vt'',\vt''') = \varphi(\vu,\vt')+\varphi(\vt',\vt'')+\varphi(\vt'',\vt''')+\varphi(\vt''',\vu),
\]
we see that the above kernel can be written as:
\[
\sum_{\vt',\vt'',\vt''' } 
e^{ib\FL(\vu,\vt',\vt'',\vt''')}s_0\kv{\vu,\vt';z-\omega_0}
j_{1,0}\kv{\vt',\vt''}
s_{0}\kv{\vt'',\vt''' ;z}
j_{2,0}\kv{\vt''' ,\vu}.
\]
It can be easily seen, using \eqref{fluxEquation}, that 
\begin{align*}
\FL &= \flux{\vu,\vt',\vt''}+\flux{\vu,\vt'',\vt'''}\\
&= \frac{1}{2} \left[
(\gu_2-\gt'_2)(\gt'_1 - \gt''_1)-(\gu_1-\gt'_1)(\gt'_2 - \gt''_2)\right]
+
\frac{1}{2} \left[(\gu_2-\gt''_2)(\gt''_1 - \gt'''_1)-(\gu_1-\gt''_1)(\gt''_2 - \gt'''_2)\right].
\end{align*}
The expansion of $e^{ib\FL}$
in $b$ is
\begin{equation}
e^{ib\FL} = 1 + ib\FL + \mathcal{O}(b^2\FL^2).\label{expAfFL}
\end{equation}
We see that due to the exponential localization of the various
kernels, the terms generated by $\mathcal{O}(b^2\FL^2)$ will give a
contribution of order $b^2$, thus it can be discarded. The linear contribution from
the right hand side of 
formula~\eqref{expAfFL} is:
\begin{align*}
&
\frac{ib}{2} \left[(\gu_2-\gt'_2)(\gt'_1 - \gt''_1)-(\gu_1-\gt'_1)(\gt'_2 -  \gt''_2) \right] 
+ \frac{ib}{2} \left[(\gu_2-\gt''_2)(\gt''_1 - \gt'''_1)-(\gu_1-\gt''_1)(\gt''_2 - \gt'''_2) \right] \\
&=
\frac{ib}{2} \left[(\gu_2-\gt'_2)(\gt'_1 - \gt''_1)-(\gu_1-\gt'_1)(\gt'_2 -\gt''_2)\right] \\
&\quad+\frac{ib}{2}\left[
(\gu_2-\gt'_2)(\gt''_1-\gt'''_1)-(\gu_1-\gt'_1)(\gt''_2-\gt'''_2) 
+(\gt'_2-\gt''_2)
(\gt''_1-\gt'''_1)- (\gt'_1-\gt''_1)(\gt''_2-\gt'''_2)
\right].
\end{align*}
Introduce this into the formula for $\left[ \szm \op{j}_{1,\subb}\sz
  \op{j}_{2,\subb}\right]\kv{\vu,\vu}$.  We now have that the linear
term in $b$ of $\left[ \szm \op{j}_{1,\subb}\sz
\op{j}_{2,\subb}\right]\kv{\vu,\vu}$ is given by:
\begin{align*}
&\frac{ib}{2} \sum_{\vt^{'},\vt^{''},\vt^{'''} \in \gT}\\
& \left[(\gu_2-\gt'_2)s_0\kv{\vu,\vt';z-\omega_0}\,(\gt'_1-\gt''_1)j_{10}\kv{\vt',\vt''}\,
s_0\kv{\vt'',\vt''';z}j_{20}\kv{\vt''',\vu} \right.\\
&- (\gu_1-\gt'_1)s_0\kv{\vu,\vt';z-\omega_0}\,(\gt'_2-\gt''_2)j_{10}\kv{\vt',\vt''}\,
s_0\kv{\vt'',\vt''';z}j_{20}\kv{\vt''',\vu} \\
&+(\gu_2-\gt'_2)s_0\kv{\vu,\vt';z-\omega_0}\,j_{10}\kv{\vt',\vt''}
\,(\gt''_1-\gt'''_1)s_0\kv{\vt'',\vt''';z}\, j_{20}\kv{vt''',\vu}\\
&-(\gu_1-\gt'_1)s_0\kv{\vu,\vt';z-\omega_0}\,j_{10}\kv{\vt',\vt''}
\,(\gt''_2-\gt'''_2)s_0\kv{\vt'',\vt''';z}\, j_{20}\kv{vt''',\vu}\\
&+s_0\kv{\vu,\vt';z-\omega_0}\,(\gt'_2-\gt''_2)j_{10}\kv{\vt',\vt''}
\,(\gt''_1-\gt'''_1)\,s_0\kv{\vt'',\vt''';z}\,j_{20}\kv{\vt''',\vu}\\
&\left. -
s_0\kv{\vu,\vt';z-\omega_0}\,(\gt'_1-\gt''_1)j_{10}\kv{\vt',\vt''}
\,(\gt''_2-\gt'''_2)\,s_0\kv{\vt'',\vt''';z}\,j_{20}\kv{\vt''',\vu}\right].
\end{align*}
Switching to $k$-space, we have that multiplying an operator-kernel
$a(\vt,\vt')$ with $(\gt_{\nu}-\gt_{\nu}')$
transfers into differentiating the fiber with respect to $k_{\nu}$, $\nu\in\{1,2\}$:
\begin{equation}
i (\gt_{\nu}-\gt_{\nu}') a(\vt,\vt') \quad\to\quad \pdiff{}{k_{\nu}} a(\vu,\vu';\v{k}). 
\label{SKType5}
\end{equation}
For example, when computing the local traces we can make the switch
\[
\sum_{\cdots\vt,\vt'\cdots \in \gT} (\cdots)j_{\nu,0}\kv{\vt,\vt'}(\cdots) \to  
\sum_{\cdots\vu,\vu'\cdots \in \Omega}\frac{1}{ \abs{\gBZ} } \int_{\gBZ} \diffd^2k\,
(\cdots)\pdiff{}{k_{\nu}} \tilde{h}_0\kv{\vu,\vu';\v{k}}(\cdots)
\]
We thus have that the coefficient of the linear term in $b$ of 
\[
\Re\frac{1}{2\abs{\gUC}\pi\omega_0} 
\oint_{\mathscr{C}} \diffd z \;f_{\subFD}(z)
\sum_{\vu \in \gUC}\left[ \szm \op{j}_{1,\subb}\sz
  \op{j}_{2,\subb}\right]\kv{\vu,\vu}   
\]
is given by (remember that $r_0\kv{z,\v{k}}$ is the matrix $(\tilde{h}_0\kv{\v{k}}-z)^{-1})$:
\begin{align}\label{decemb3}
  &\Re\left(\frac{i}{4\abs{\gUC}\pi\omega_0\abs{\gBZ}} 
  \oint_{\mathscr{C}} \diffd z \;f_{\subFD}(z)\int_{\gBZ}
  \diffd^2k\right.\sum_{\vu,\ldots,\vu''' \in \gUC}\nonumber \\
&\left[
-\pdiff{}{k_2}r_0\kv{\vu,\vu';z-\omega_0,\v{k}}\frac{\partial^2}{\partial
  k_1^2}\tilde{h}_0\kv{\vu',\vu'';\v{k}} r_0\kv{\vu'',\vu''';z,\v{k}} \pdiff{}{k_2}\tilde{h}_0\kv{\vu''',\vu;\v{k}}\right.\nonumber\\
&+\pdiff{}{k_1}
r_0\kv{\vu,\vu';z-\omega_0,\v{k}}\frac{\partial^2}{\partial k_2 \partial k_1}\tilde{h}_0\kv{\vu',\vu'';\v{k}}
r_0\kv{\vu'',\vu''';z,\v{k}} \pdiff{}{k_2}\tilde{h}_0\kv{\vu''',\vu;\v{k}}\nonumber\\
&-\pdiff{}{k_2}r_0\kv{\vu,\vu';z-\omega_0,\v{k}}\pdiff{}{k_1}\tilde{h}_0\kv{\vu',\vu'';\v{k}}
\pdiff{}{k_1}r_0\kv{\vu'',\vu''';z,\v{k}}\pdiff{}{k_2}\tilde{h}_0\kv{\vu''',\vu;\v{k}}\nonumber
\\
&+\pdiff{}{k_1}r_0\kv{\vu,\vu';z-\omega_0,\v{k}}\pdiff{}{k_1}\tilde{h}_0\kv{\vu',\vu'';\v{k}}
\pdiff{}{k_2}r_0\kv{\vu'',\vu''';z,\v{k}}\pdiff{}{k_2}\tilde{h}_0\kv{\vu''',\vu;\v{k}}\nonumber
\\
&-r_0\kv{\vu,\vu';z-\omega_0,\v{k}} \frac{\partial^2}{\partial
  k_2\partial k_1}\tilde{h}_0\kv{\vu',\vu'';\v{k}}
 \pdiff{}{k_1}r_0\kv{\vu'',\vu''';z,\v{k}} \pdiff{}{k_2} \tilde{h}_0 \kv{\vu''',\vu;\v{k}}\nonumber\\
&\left.\left.+r_0\kv{\vu,\vu';z-\omega_0,\v{k}}  \frac{\partial^2}{\partial
  k_1^2} \tilde{h}_0\kv{\vu',\vu'';\v{k}}
\pdiff{}{k_2}r_0\kv{\vu'',\vu''';z,\v{k}}\pdiff{}{k_2}\tilde{h}_0\kv{\vu''',\vu;\v{k}}\right]\right).
\end{align}
Now consider the factor $\sk = \op{S}_{\subb}(z)\op{K}_{\subb}(z)$ which is a  ``SK-type'' operator:
\begin{equation}
  \label{SKtype}
  \begin{split}
&\left[
\op{S}_{\subb}(z)\op{K}_{\subb}(z)
\right]\kv{\vt,\vtt} \\
&=
\sum_{\vt',\vt'' \in \gT} e^{ib\varphi(\vt,\vt')}s_0\kv{\vt,\vt';z}
 e^{ib\varphi(\vt',\vtt)}\left( e^{ib\flux{\vt',\vt'',\vtt}} -1\right) h_0\kv{\vt',\vt''} s_0\kv{\vt'',\vtt;z}.
\end{split}
\end{equation}
We see that $e^{ib\flux{\vt',\vt'',\vtt}} -1$ is already first order
in $b$, thus we can discard the two terms $\skm\op{j}_{1b} \sk \op{j}_{2b}$ (one from $z$ and one from $z \to z+\omega_0)$
 in \eqref{hejsa}, when neglecting all terms not linear in $b$. Furthermore, we can reduce \eqref{SKtype} to  
\begin{equation}
e^{ib\varphi(\vt,\vtt)}\sum_{\vt',\vt'' \in \gT}s_0\kv{\vt,\vt';z}
 \left( e^{ib\flux{\vt',\vt'',\vtt}} -1\right) h_0\kv{\vt',\vt''} s_0\kv{\vt'',\vtt;z},
 \label{SKtype2}
 \end{equation}
when neglecting higher orders of $b$. The first order contribution of $e^{ib\flux{\vt',\vt'',\vtt}} -1$ is
\begin{equation}
\frac{1}{2}
ib\flux{\vt',\vt'',\vtt} = \frac{ib}{2}
\left[(\gt_2'-\gt_2'')(\gt_1''-\gtt_1)-(\gt_1'-\gt_1'')(\gt_2''-\gtt_2)\right].
\label{SKtype3}
\end{equation}
In first order in $b$, \eqref{SKtype} gives:
\begin{equation}
\frac{ib}{2}\sum_{\vt',\vt'' \in \gT} s_0\kv{\vt,\vt';z}
\left[i(\gt_1'-\gt_1'')i(\gt_2''-\gtt_2)-i(\gt_2'-\gt_2'')i(\gt_1''-\gtt_1)\right]
h_0\kv{\vt',\vt''} s_0\kv{\vt'',\vtt;z},
\label{SKtype4}
\end{equation}
where we have inserted some $i$'s to bring it on the form \eqref{SKType5}.
Using~\eqref{SKtype4} in~
expression~\eqref{SKtype}, we have that the linear contribution of 
\[
\Re\frac{1}{2\abs{\gUC}\pi\omega_0} 
\oint_{\mathscr{C}} \diffd z \;f_{\subFD}(z)\sum_{\vu \in \gUC}
  \left[\szm \op{j}_{1b} \sk\op{j}_{2b} \right](\vu,\vu)
\]
  is:
\begin{align} 
 & \Re\frac{ib}{4\abs{\gUC}\pi\omega_0\abs{\gBZ}} 
\oint_{\mathscr{C}} \diffd z f_{\subFD}(z)\int_{\gBZ}
\diffd^2k\nonumber \\
&\sum_{\vu \in \gUC}\left[
  r_0(z-\omega_0)\, \pa_1 \tilde{h}_0\, r_0(z)
  [\pa_1 \tilde{h}_0 \, \pa_2r_0(z) - \pa_2\tilde{h}_0\, \pa_1 r_0(z)]
  \pa_2\tilde{h}_0
  \right](\vu,\vu),
  \label{SKtype6}
\end{align}
where we have suppressed the $\v{k}$-dependence of the operators, and use the notation $\pa_1=\pdiff{}{k_1}$ and $\pa_1=\pdiff{}{k_2}$.

\subsubsection{All terms in $k$-space}
Using the method above, we can calculate the traces of
\eqref{hejsa}, for a given
$z,z\pm \omega$ in the resolvent set of $\op{H}_0$, explicitly, by simply
inverting and differentiating known $\abs{\gUC}\times \abs{\gUC}$-matrices.

Written in $k$-space, these terms are:
\begin{align*}
 & \sum_{\vu \in \gUC} \left[\szm\op{j}_{1b}\sz \op{j}_{2b}\right](\vu,\vu): 
\\
&\frac{ib}{2\abs{\gBZ}}\int_{\gBZ} \diffd^2 k\,\sum_{\vu \in \gUC} \left[
\left(\pa_1 r_0(z-\omega_0)\pa_2\pa_1\tilde{h}_0 
     -\pa_2 r_0(z-\omega_0)\pa_1^2   \tilde{h}_0  \right)r_0(z)\pa_2\tilde{h}_0\right.\\
&\qquad+\left(\pa_1 r_0(z-\omega_0) \pa_1 \tilde{h}_0 \pa_2 r_0(z)-
        \pa_2 r_0(z-\omega_0) \pa_1 \tilde{h}_0 \pa_1 r_0(z)\right) \pa_2 \tilde{h}_0\\
&\qquad+\left.r_0(z-\omega_0)\left(\pa_1^2    \tilde{h}_0 \pa_2 r_0(z) -
	\pa_2 \pa_1\tilde{h}_0 \pa_1 r_0(z)\right)\pa_2\tilde{h}_0
\right](\vu,\vu).
\end{align*}
\begin{align*}
 & \sum_{\vu \in \gUC} \left[\szm\op{j}_{1b}\sk \op{j}_{2b}\right](\vu,\vu): 
\\
&\frac{ib}{2\abs{\gBZ}}\int_{\gBZ} \diffd^2 k\,\sum_{\vu \in \gUC} \left[
r_0(z-\omega_0)\pa_1\tilde{h}_0r_0(z)
\left( \pa_1\tilde{h}_0\pa_2r_0(z)-\pa_2\tilde{h}_0\pa_1 r_0(z)\right)\pa_2\tilde{h}_0
\right](\vu,\vu).
\end{align*}
\begin{align*}
 & \sum_{\vu \in \gUC} \left[\skm\op{j}_{1b}\sz \op{j}_{2b}\right](\vu,\vu): 
\\
&\frac{ib}{2\abs{\gBZ}}\int_{\gBZ} \diffd^2 k\,\sum_{\vu \in \gUC} \left[
r_0(z-\omega_0)
\left( \pa_1\tilde{h}_0\pa_2r_0(z-\omega_0)-\pa_2\tilde{h}_0\pa_1 r_0(z-\omega_0)\right)
\pa_1\tilde{h}_0r_0(z)\pa_2\tilde{h}_0
\right](\vu,\vu).
\end{align*}
\subsubsection{Collecting the terms}
Using that $\abs{\gUC}\abs{\gUC^*}=4\pi^2$ and
inserting everyting into formula~\eqref{bConduction2}
we have:
\begin{equation}
\label{firstDerivativeMonsterFormula}
\begin{split}
&\sigma_{21}^{(1)}=\frac{1}{16\pi^3\omega_0} \int_{\gBZ} \diffd^2 k\,\Re
\oint_{\mathscr{C}} \diffd z \;if_{\subFD}(z)\sum_{\vu \in \gUC}\\
&\quad\left[ \left( \pa_1 r_0(z-\omega_0)\pa_2\pa_1\tilde{h}_0 
        -\pa_2 r_0(z-\omega_0)\pa_1^2   \tilde{h}_0  \right) r_0(z)\pa_2\tilde{h}_0\right.\\
&\quad+\left(\pa_1 r_0(z-\omega_0) \pa_1 \tilde{h}_0 \pa_2 r_0(z)-
        \pa_2 r_0(z-\omega_0) \pa_1 \tilde{h}_0 \pa_1 r_0(z)\right) \pa_2 \tilde{h}_0\\
&\quad+r_0(z-\omega_0)\left(\pa_1^2    \tilde{h}_0 \pa_2 r_0(z) -
	\pa_2 \pa_1\tilde{h}_0 \pa_1 r_0(z)\right)\pa_2\tilde{h}_0\\
&\quad+
r_0(z-\omega_0)\pa_1\tilde{h}_0r_0(z)
\left( \pa_1\tilde{h}_0\pa_2r_0(z)-\pa_2\tilde{h}_0\pa_1 r_0(z)\right)\pa_2\tilde{h}_0\\
&\quad+
r_0(z-\omega_0)
\left( \pa_1\tilde{h}_0\pa_2r_0(z-\omega_0)-\pa_2\tilde{h}_0\pa_1 r_0(z-\omega_0)\right)
\pa_1\tilde{h}_0r_0(z)\pa_2\tilde{h}_0\\
&\quad+ \left.\{z\to z+\omega_0\}\right](\vu,\vu).
\end{split}
\end{equation}
This formula only contains known matrices and their derivatives. An example of a numerical investigation of this formula, used to calculate the optical Hall conductivity in a nearest neighbour tight-binding model of graphene, is given in \cite{PhysRevB.86.235438}.

\subsection{Consequences of the symmetry}
Now we want to prove \eqref{TaylorCoefficient}. The
following lemmas will help us prove that all even Taylor
coefficients of $\sigma_{21}^G(b)$ vanish.
\begin{lemma}
  \label{XLemma1}
  For any $n+1$-tuple of sites $( \vt, \vt^{(1)}, \ldots , \vt^{(n)} )$ in $\gT$, it holds that
  \begin{equation}
   \varphi(\vt,\vt^{(1)}) + 
   \sum_{m=1}^{n-1} \varphi(\vt^{(m)} , \vt^{(m+1)})+
   \varphi(\vt^{(n)} , \vt) 
 = \sum_{m=1}^{n-1}\flux{\vt, \vt^{(m)},\vt^{(m+1)}}.
   \label{OddLemma1}
  \end{equation}
  \begin{proof}
    By telescoping, the left hand
    side of the above equation can be rewritten as
  $$
   \sum_{m=1}^{n-1}\left [  \varphi(\vt, \vt^{(m)}) +\varphi(\vt^{(m)}, \vt^{(m+1)}) +
   \varphi(\vt^{(m+1)}, \vt)\right ].
 $$
 Identity \eqref{fluxEquation} and the anti-symmetry of $\varphi$ give \eqref{OddLemma1}.
 \end{proof}
\end{lemma}
\begin{lemma}
   \label{XLemma2}
   Given any $(n+1)$-tuple of sites $( \vt, \vt^{(1)}, \ldots
   ,\vt^{(n)} )$ in $\gT$, and an index $1 \leq r \leq n-1$,
   it holds that $\flux{\vt, \vt^{(r)}, \vt^{(r+1)}}$ is given by
   \begin{align}\label{decemb4}
   \frac{1}{2} \left[ 
   (\vt^{(r)} - \vt^{(r+1)})\times(\vt - \vt^{(1)}) 
+ \sum_{m=1}^{r-1} (\vt^{(r)} - \vt^{(r+1)})\times(\vt^{(m)} - \vt^{(m+1)})\right]_z.
  \end{align}
   \begin{proof}
     A telescoping argument gives that
     \begin{align}
       \flux{\vt, \vt^{(r)}, \vt^{(r+1)}} &= \frac{1}{2} \left[
	(\vt^{(r)} - \vt^{(r+1)})\times(\vt - \vt^{(r)})
	\right]_z \nonumber \\
	&= \frac{1}{2} \left[ 
	(\vt^{(r)} - \vt^{(r+1)})\times
	\left(\vt - \vt^{(1)}+\sum_{m=1}^{r-1}
	(\vt^{(m)} - \vt^{(m+1)})\right)
	\right]_z, \nonumber
      \end{align}
      which proves the lemma.
   \end{proof}
   \end{lemma}

\vspace{0.5cm}

   Suppose that we want to determine the $n$'th Taylor coefficient ($n\geq2$) given like  \eqref{TaylorCoefficient}.
   For $z \in \rho (\op{H}_{\subb})$ put $\tilde{z}=z-\omega_0$. The problem is basically to 
   identify the coefficient to $b^n$ in matrix elements of the type
   \begin{equation}
	 \left[
	 \left(\sum_{r=0}^{n} 
	 \op{S}_{\subb}(\tilde{z})\left(\op{K}_{\subb}(\tilde{z})\right)^r\right)
	\op{j}_{1,\subb}
	\left(\sum_{r=0}^{n} 
	 \op{S}_{\subb}(z)\left(\op{K}_{\subb}(z)\right)^r\right)
	 \op{j}_{2,\subb} \right](\vu,\vu),	
	 \label{genericTC}
   \end{equation}
   as we did for $n=1$ in expression \eqref{bConduction2}. Expression \eqref{genericTC} can be expanded into a finite sum of terms all in the form $\left[\op{S}_{\subb}(\tilde{z})\left(\op{K}_{\subb}(\tilde{z})\right)^{q_1}
   \op{j}_{1,\subb}\op{S}_{\subb}(z)\left(\op{K}_{\subb}(z)\right)^{q_2}\op{j}_{2,\subb}\right](\vu,\vu)$ for
   some $0\leq q_1, q_2 \leq n$.  Expanding one such term as a sum
   over products of integral kernels we obtain (to shorten notation, we write from
   now on $\vt^{m}$ instead of $\vt^{(m)}$):
   \[
   \left[\op{S}_{\subb}(\tilde{z})\left(\op{K}_{\subb}(\tilde{z})\right)^{q_1}
   \op{j}_{1,\subb}\op{S}_{\subb}(z)\left(\op{K}_{\subb}(z)\right)^{q_2}\op{j}_{2,\subb}\right](\vu,\vu)
   =
   \sum_{\vt^{i} \in \gT, i=1\ldots q} F_1(\vu,\vt^1)F_2(\vt^1,\vt^2)\ldots F_q(\vt^q,\vu),
   \]
   where each $F_i(\cdot,\cdot)$ is an operator kernel of either an
   $\op{S}$, a $\op{K}$ or a $\op{j}$ operator.
   The $b$-dependence of such an expression is always given in the form
   \begin{equation}
     M(b) = e^{ib[\varphi(\vu,\vt^1)+\ldots+  \varphi(\vt^{i},
       \vt^{i+1})+\ldots  + \varphi(\vt^{q_3}, \vu)]}\tilde{M}(b), \label{bDep}
   \end{equation}
   where $\tilde{M}(b)$ is given by a convolution of kernels at
   $b=0$, together with factors of the type
   $
   \left(e^{ib\flux{\vtt,\vtt',\vtt'''}}-1\right)$, 
   where $\vtt$, $\vtt''$ and $\vtt'''$ are consecutive convolution
   variables. Together with lemmas \ref{XLemma1} and 
   \ref{XLemma2}, it follows that the phases can be expanded in such a
   way that the only type of factors which can appear are of the
   form 
  \[ \alpha(\vt, \vt', \vt'', \vt''')= \left[ (\vt - \vt')\times
	(\vt'' - \vt''')
	\right]_z=(x_1-x_1')(x_2''-x_2''')-(x_1''-x_1''')(x_2-x_2'),
   \]
where $\vt$ and $\vt'$ (respectively $\vt''$ and $\vt'''$) are
consecutive variables in the convolutions. 

The coefficient of $b^n$ will consist of a finite number of
convolutions, each of which having $n$ such $\alpha$ factors. They
will generate $n$ factors of type $y_2-y_2'$, where $\vtt$ and $\vtt'$
are consecutive convolution variables. 

We have to keep in mind that $\sigma_{21}$ contains from the beginning
an $x_2-x_2'$ coming from $\op{j}_{2}$. Thus the convolutions giving
$b^n$ will all have exactly $n+1$ factors like $x_2-x_2'$, where $\vt$ and $\vt'$
are consecutive convolution variables. Switching to the $k$ space,
these factors will be transformed into $n+1$ partial derivatives with
respect to $k_2$. Remember that all matrix elements of
$\tilde{h}_0^G({\bf k})$ and $r_0^G(z,{\bf k})$ are even functions of $k_2$. Distributing $2p+1$ 
derivatives with respect to $k_2$ among these matrix
elements will generate a global odd function in $k_2$. When
integrating with respect to $k_2$ over the symmetric first Brillouin
zone, we get  zero. Thus \eqref{TaylorCoefficient} is proved.

\section{Conclusions}\label{unu4}

\begin{enumerate}
\item We constructed the conductivity tensor going through the
  Kubo-Greenwood formalism, paying attention to the thermodynamic and
  adiabatic limits even though most physical papers completely ignore these
  issues. Our proof of the
  thermodynamic limit is 
  based on a simplified version of the geometric perturbation theory as developed in
  \cite{BCD} for the Schr\"odinger case and then further developed in
  \cite{CN}. 
\item The gauge-invariant magnetic perturbation theory cannot be avoided if one wants
  to control the growth at infinity induced by the constant magnetic
  field. Moreover, it provides us with a systematic method of
  computing derivatives of any order at $b=0$. 
\item Remember that the Faraday rotation
$\vartheta$ is proportional with $\sigma_{21}'(0)$ (denoted by
$\sigma_{21}^{(1)}$ in formula \eqref{FaradaySigma}, which gives the Verdet constant). 
In \eqref{firstDerivativeMonsterFormula} we obtain a closed formula
for $\sigma_{21}^{(1)}$ which will be the starting point of a further
analysis of the dependency of the Verdet constant on temperature,
chemical potential, density and spectral structure of a given
material.For instance, the graphene is very interesting because at zero
temperature the chemical potential lies exactly where the valence and
conduction energy bands touch each other (see points $K$ and $K'$ in figure
\ref{fig:bandstructure}). The eventual lack of regularity of the Fermi
surface can make the zero temperature limit nontrivial. 
Formula has subsequently been used to calculate the optical Hall conductivity 
in a nearest neighbour tight-binding model of graphene, see \cite{PhysRevB.86.235438}.

\item The expression giving $\sigma_{21}^{(1)}$
has an analytic extension in $\omega_0$ to the whole complex plane
except in zero and those real values for which the sets $\sigma(\op{H}_0)\pm
\omega_0$ have common points with  $\sigma(\op{H}_0)$. But for
certain particular models one can further extend the permitted regions of
$\omega_0$. In fact, it would be very interesting to study how 
$\sigma_{21}^{(1)}$ behaves when $\omega_0$ comes close
to those resonant values which induce transitions between different
Bloch-bands. 

\item In the case when the magnetic field $b$ generates a rational flux through
  the unit cell of $\op{H}_0$, then the spectrum of $\op{H}_b$ consists of bands, but the elementary cells  
  of $\op{H}_b$ become
  larger and larger when $b$ becomes smaller and smaller. Nevertheless, one can
  compute $\sigma_{21}(b)$ in terms of the $b$-dependent Bloch
  structure, see e.g. \cite{PhysRevB.84.115424}. We 
  already compared this approach with our method in \cite{PhysRevB.86.235438}. 
  The results are almost identical, even though the computational effort implied by our method is considerably lower.

\item An open problem: take our graphene Hamiltonian $\op{H}_0^G$
  whose kernel is given in
  \eqref{decemb10}, and put a weak magnetic field on it through a Peierls
  phase. What happens with the spectrum of $\op{H}_b^G$ around the
  crossing of the valence and conduction bands, represented in figure
  \ref{fig:bandstructure}? Physicists claim that in that energy region
  the dynamics is close to the one generated by some zero-mass Dirac operator, and when we
  add a magnetic field it should create gaps which behave like $\sqrt{b}$.   

\item We note that the first term of \eqref{mainResult1}
disappears only after the adiabatic limit ($\eta=0$). For graphene,
unlike the usual Schr\"odinger operators, the commutator $[\op{j}_{2,\subb,\subN}, \op{X}_{1,\subN}
]$ is not zero. 
\end{enumerate}

\section{Acknowledgments}
The authors acknowledge partial support from the Danish FNU grant {\it Mathematical Analysis of Many-Body
Quantum Systems}. Part of this work was done at Institut Mittag-Leffler during the program {\it Hamiltonians
in Magnetic Fields}.

%

\end{document}